\documentclass[12pt]{article}
\pdfoutput=1
\usepackage{amsmath,amsfonts,graphicx,color,bbm,tikz,bm,setspace}
\usepackage{comment}
\usepackage[nosort]{cite}
\usepackage{subfigure}
\usetikzlibrary{calc,positioning}
\usetikzlibrary{patterns,arrows,decorations.pathreplacing}
\usepackage{caption}
\usepackage{ulem}
\tikzset{>=stealth}
\usepackage{lipsum}
\usepackage{tabularx}
\usepackage{fullpage}
\usepackage{hyperref}
\usepackage{bbold}
\usepackage{amsthm}
\theoremstyle{definition}
\newtheorem{definition}{Definition}
\newtheorem{theorem}{Theorem}
\usepackage{listings}
\usepackage{tcolorbox}

\newcommand{\mathsym}[1]{{}}
\newcommand{\unicode}[1]{{}}

\makeatletter\@addtoreset{equation}{section}\makeatother

\newcommand{\be}{\begin{equation}}
\newcommand{\ee}{\end{equation}}
\def\beq{\begin{equation}}
\def\eeq{\end{equation}}
\newcommand{\bea}{\begin{eqnarray}}
\newcommand{\eea}{\end{eqnarray}}

\renewcommand{\title}[1]{\vbox{\center\LARGE{#1}}\vspace{3mm}}
\renewcommand{\author}[1]{\vbox{\center{#1}}\vspace{3mm}}

\newcommand{\email}[1]{\vbox{\center\tt#1}\vspace{3mm}}

\begin{document}
\begin{titlepage}

\begin{center}
{\large {\bf Solving the Yang-Baxter, tetrahedron and higher simplex equations using Clifford algebras}}

\author{ Pramod Padmanabhan,$^a$
and Vladimir Korepin$^{b}$}

{$^a${\it School of Basic Sciences,\\ Indian Institute of Technology, Bhubaneswar, 752050, India}}
\vskip0.1cm
{$^b${\it C. N. Yang Institute for Theoretical Physics, \\ Stony Brook University, New York 11794, USA}}

\email{pramod23phys@gmail.com, vladimir.korepin@stonybrook.edu}

\vskip 0.5cm 

\end{center}


\abstract{
\noindent 
Bethe Ansatz was discoverd in 1932. Half a century later its algebraic structure  was unearthed: Yang-Baxter equation was discovered, as well as its multidimensional generalizations [tetrahedron equation and $d$-simplex equations]. Here we describe a universal method to solve these equations using Clifford algebras. The Yang-Baxter equation ($d=2$), Zamalodchikov's tetrahedron equation ($d=3$) and the Bazhanov-Stroganov equation ($d=4$) are special cases. Our solutions form a linear space. This helps us to include spectral parameters. Potential applications are discussed. 
}

\end{titlepage}
\tableofcontents 

\section{Introduction}
\label{sec:Introduction}
Quantum integrable models in 1+1 dimensions are described and solved using the Yang-Baxter equation \cite{YangCN1967, BAXTER1972193} and the associated quantum inverse scattering \cite{takhtadzhyan1979,Takhtadzhan_1979}  or algebraic Bethe ansatz method \cite{Korepin1993QuantumIS}. The Yang-Baxter equation appears in several contexts : in the theory of quantum groups \cite{Majid_1995}, knot theory and braid groups \cite{Turaev1988TheYE, Kauffman1991}, in exactly solvable models in statistical mechanics \cite{Baxter1982ExactlySM} and in factorizable scattering theory \cite{Zamolodchikov1979FactorizedSI}. An instinctive question is the extension of these ideas to higher dimensions. The first steps in this direction were taken by Zamalodchikov in introducing the three dimensional version known as the tetrahedron equation \cite{Zamolodchikov1980TetrahedronOriginal,Zamolodchikov1981TetrahedronEA}. This equation describes the scattering of straight strings in 2+1 spacetime dimensions. The initial years afters its introduction saw a flurry of papers studying various aspects of the model starting from its solutions and corresponding three dimensional integrable models. Examples of the latter are found in \cite{Bazhanov1984FreeFO,Bazhanov1992NewSL,Bazhanov1993StartriangleRF,Stroganov1997TetrahedronEA,Korepanov1989TETRAHEDRALZA,Sergeev1995TheVF,Mangazeev2013AnI3,Talalaev2015ZamolodchikovTE,Khachatryan2015IntegrabilityIT,Maillet1989IntegrabilityFM,Kashaev_1998}. The first non-trivial solutions to the tetrahedron equation appeared in \cite{Zamolodchikov1980TetrahedronOriginal} which was further explained in \cite{Baxter1983OnZS}. Subsequently several other solutions have been constructed \cite{Korepanov1993TetrahedralZA,Bazhanov2005ZamolodchikovsTE,Hietarinta_1994,Korepanov1994TheTE,Bazhanov2009QuantumGO}. Solutions can be obtained from cluster algebras\cite{sun2022cluster,Gavrylenko2020SolutionOT,inoue2024tetrahedron, inoue2023quantum,inoue2024solutions}, and quantum groups \cite{Kuniba2015TetrahedronEA}. Many of the solutions in these works are obtained from the so called Tetrahedron Zamalodchikov algebra (TZA)\footnote{See \cite{korepanov2013novel} for a short description and \cite{WadatiShiroishi1995, WadatiShiroishi1995(2)} for its role in the integrability of the 1D Hubbard model.} which is the relation 
\begin{equation}\label{eq:TZA}
    R^a_{12}R^b_{13}R^c_{23} = \sum\limits_{d,e,f=0}^1~\left(S_{123} \right)_{def}^{abc}R_{23}^fR_{13}^eR_{12}^d.
\end{equation}
The $S$ operator in this equation satisfies the tetrahedron equation in \eqref{eq:tetrahedronEdgeForm}. More solutions obtained from $R$-matrices include \cite{Kuniba2015CombinatorialYM,Kuniba2018TetrahedronEA}. Along these lines reflection equations for the tetrahedron case can also be solved \cite{Kuniba2012TetrahedronA3,Kuniba2018MatrixPS,Isaev_1997}. As a result solving the tetrahedron equation helps us obtain an infinite number of Yang-Baxter solutions \cite{Bazhanov2009QuantumGO}. The higher ($d\geq 4$) simplex equations first appeared in \cite{Bazhanov1982ConditionsOC,MAILLET1989221}. However the amount of work surrounding these equations is far lesser than the tetrahedron equation. Some solutions can be found in \cite{Hietarinta_1997,CarterSaito,Frenkel1991SimplexEA,bardakov2022settheoretical}. The general philosophy in these works is also to use the $d-1$-simplex operators to generate the solutions of the $d$-simplex equations. 

In this paper we will use an entirely different approach that is completely algebraic.  This method is universal in the sense that it provides the algorithm to construct the $d$-simplex operators for all the $d$. The starting point is a pair of anticommuting operators $A$ and $B$ that act on the local Hilbert space. First we write down the simplest solutions of the constant [independent of spectral parameter] $d$-simplex equations. Next we show that these solutions form a linear space. The coefficients in their linear combinations are interpreted as spectral parameters. Thus we also solve the spectral parameter dependent $d$-simplex equations. For easier navigation we summarise the main results.

\subsection*{Summary of main results}
\label{subsec:summaryIntroduction}
The following summary is intended for the reader who wishes to go directly to the solutions of the various $d$-simplex operators. A key feature of our solutions is that they form a linear space. All solutions are representation independent. 
\begin{enumerate}
    \item {\it $d=2$ or {\bf Yang-Baxter operators} :} Solutions in \eqref{eq:Rd2} and \eqref{eq:Rd2general}.
    \item {\it $d=3$ or {\bf Tetrahedron operators} :} Solutions in \eqref{eq:Rd3}, \eqref{eq:Rd3general} and \eqref{eq:Rd3generalBA}.
    \item {\bf 4-simplex operators :} Solutions in \eqref{eq:R1d4}, \eqref{eq:R1d4general} and \eqref{eq:R2d4}.
    \item {\bf 5-simplex operators :} Solutions in \eqref{eq:R1d5} and \eqref{eq:R2d5}.
    \item The method to write down the answers for a general $d$ are captured in Theorems 1 and 2 in Sec. \ref{subsec:gend}.
    \item The qubit versions can be found in Sec. \ref{sec:qubitsolutions}.
\end{enumerate}

\subsection*{Organisation of the paper}
\label{subsec:organisation}
There is no unique form for the higher simplex equations. The various forms and index structures are discussed in Sec. \ref{sec:labeling}. We solve the vertex forms of the higher simplex equations. Starting from $d=2$ we present the method to solve all the $d$-simplex equations using Clifford algebras in Sec. \ref{sec:solutions}. The solutions in this section are algebraic, meaning they are independent of the choice of representation of the local Hilbert space. For each choice we get a different set of solutions. The linear structure of the space of solutions is discussed for each $d$. The operators obtained by choosing the qubit representation is shown in Sec. \ref{sec:qubitsolutions}. We conclude with a summary of all the solutions and future directions in Sec. \ref{sec:conclusion}.

We include several important appendices that present the versatility of the method developed in the main text. 
\begin{enumerate}
    \item Appendix \ref{app:ABalphad3} is a technical section which shows  that the $A$ and $B$ operators need to either commute or anticommute to solve the $d$-simplex equations for $d\geq 3$.
    \item In Appendix \ref{app:antidsimplex} we define and solve the anti-$d$-simplex equation. This is the equation where the right hand side is multiplied by an overall negative sign when $d$ is even. The odd case is also discussed. This naturally occurs in this construction.
    \item The tetrahedron equation takes different forms. We show that the Clifford algebra method provides a solution for each of these forms in Appendix \ref{app:othertetrahedra}.
    \item Examples of solutions to the higher reflection equations using Clifford algebras are shown in Appendix \ref{app:reflection}.
    \item The methodology to generate solutions extends to non-Clifford solutions as well. This is the subject of the last Appendix \ref{app:nonClifford}.
    \item The Mathematica codes to verify the $d$-simplex solutions in the qubit representation are shown in Appendix \ref{app:Mcodes}.
\end{enumerate}

\section{Labeling schemes}
\label{sec:labeling}
Our first task is to write down the higher simplex equations. The $d$-simplex equations describe the scattering of extended ($d-2$ dimensional) objects in $(d-1) + 1$ dimensions. Corresponding to the simplices generated in these processes there are different ways of labelling them. This is reflected in the index structure of these equations as we shall now see. We will closely follow \cite{MAILLET1989221,Hietarinta_1994,Hietarinta1997TheET} to establish the conventions.

We begin with the first non-trivial simplex equation, the tetrahedron equation. This equation describes the scattering of straight strings or particles at the intersection of strings. It was first formulated and solved by Zamalodchikov in \cite{Zamolodchikov1980TetrahedronOriginal,Zamolodchikov1981TetrahedronEA}, with subsequent simplifications and solutions \cite{Baxter1983OnZS, Bazhanov1984FreeFO}. There are multiple forms of the tetrahedron equation. The source of this ambiguity lies in the ordering of particles in two dimensions. This is visibly absent for the Yang-Baxter equation (star-triangle relations) in one dimension where an analogous ordering takes place without ambiguity. The basic object is the scattering matrix of three particles. We will denote it by $$ R^{k_1k_2k_3}_{j_1j_2j_3}$$
with the $j$'s ($k$'s) indexing the incoming (outgoing) particles respectively. The tetrahedron equation describes the scattering of four strings with six intersection points. The two ways of scattering six particles should yield the same result leading to the consistency condition described by the tetrahedron equation\footnote{The scattering process can be depicted pictorially. As the considerations in this paper are algebraic we direct the interested reader to \cite{Hietarinta_1994} for these pictures.} :
\begin{equation}
R^{k_1k_2k_3}_{j_1j_2j_3}R^{l_1k_4k_5}_{k_1j_4j_5}R^{l_2l_4k_6}_{k_2k_4j_6}R^{l_3l_5l_6}_{k_3k_5k_6} = R^{k_3k_5k_6}_{j_3j_5j_6}R^{k_2k_4l_6}_{j_2j_4k_6}R^{k_2l_4l_5}_{j_1k_4k_5}R^{l_1l_2l_3}_{k_1k_2k_3} 
\end{equation}
The $k$ indices are summed over using the Einstein summation convention. In operator form the above equation is written as :
\begin{equation}\label{eq:tetrahedronVertexForm}
    R_{123}R_{145}R_{246}R_{356} = R_{356}R_{246}R_{145}R_{123}.
\end{equation}
This is also known as the vertex-form of the tetrahedron equation. We expect six indices in this equation as they correspond to the six particles formed at the intersection of the four strings. Thus this equation acts on $\bigotimes\limits_{j=1}^6~V_j$, with $V$ being the local Hilbert space. It is natural to label the scattering matrix with the string segments of the four scattering strings. This leads to the next form of the tetrahedron equation :
\begin{equation}\label{eq:tetrahedronEdgeForm}
    R_{123}R_{124}R_{134}R_{234} = R_{234}R_{134}R_{124}R_{123}.
\end{equation}
It acts on $\bigotimes\limits_{j=1}^4~V_j$ corresponding to the four scattering strings. Finally we have the vacuum or cell labelling where the equation is given in terms of Boltzmann weights. We will omit writing down this expression\footnote{The resulting equation describes the classical three dimensional integrable model. Since that is not the subject of this work we will skip this form.} and refer the reader to \cite{Hietarinta_1994} for the explicit form. 

While the above three forms are inspired by physical scattering processes, the last form we will write down is derived from algebraic considerations. Let us assume the Yang-Baxter equation to be weakly true, that is it is true up to a conjugation. The `conjugator' is a three indexed object satisfying a consistency condition which gives another form of the tetrahedron equation\footnote{This form is equivalent to the vertex form in \eqref{eq:tetrahedronVertexForm} by interchanging the indices 3 and 4. See Appendix \ref{app:othertetrahedra} for their solutions.} :
\begin{equation}\label{eq:tetrahedronYBconjugator}
    R_{124}R_{135}R_{236}R_{456} = R_{456}R_{236}R_{135}R_{124}.
\end{equation}
This is also known as the quantized Yang-Baxter equation in \cite{kuniba2022quantum}. This consistency equation can be written down in eight different ways arising due to the ambiguity in the reversal process \cite{Hietarinta1997TheET}. This is extensively studied in the higher category and higher braid group literature \cite{Baez1995HigherDA, Kapranov19942CategoriesAZ}. In this work we will study the vertex form of the tetrahedron  equation \eqref{eq:tetrahedronVertexForm} in the main text and reserve the forms in \eqref{eq:tetrahedronEdgeForm} and \eqref{eq:tetrahedronYBconjugator} for Appendix \ref{app:othertetrahedra}.

The next higher simplex equation is the 4-simplex equation. This is also known as the Bazhanov-Stroganov equation \cite{Bazhanov1982ConditionsOC}. The vertex form of this equation reads :
\begin{equation}\label{eq:4simplexVertexForm}
    R_{1234}R_{1567}R_{2589}R_{368,10}R_{479,10} 
     =  R_{479,10}R_{368,10}R_{2589}R_{1567}R_{1234}.
\end{equation}
The index structure of the 4-simplex equation suggests the pattern for the vertex form of the higher simplex equations \cite{MAILLET1989221}. For example the vertex form of the 5-simplex equation is 
\begin{eqnarray}\label{eq:5simplexVertexForm}
    & & R_{12345}R_{16789}R_{26,10,11,12}R_{37,10,13,14}R_{48,11,13,15}R_{59,12,14,15} \nonumber \\
    & = & R_{59,12,14,15}R_{48,11,13,15}R_{37,10,13,14}R_{26,10,11,12}R_{16789}R_{12345}. 
\end{eqnarray}
For the sake of completion we write the algorithm to construct the vertex form of the $d$-simplex equation for arbitrary $d$. The $d$-simplex operator acts non-trivially on $d$ sites. On each side of the equation there are a product of $d+1$, $d$-simplex operators. They act on $$\bigotimes\limits_{j=1}^{\frac{d(d+1)}{2}}~V_j,$$ with $V$ being the local Hilbert space. The number $\frac{d(d+1)}{2}$ corresponds to the number of particles formed at the vertices of the scattering process of $d+1$ extended objects. Now we construct the left hand side of the $d$-simplex equation. Begin with the operator $R_{1,2\cdots, d}$. The second operator starts with the index 1 and acquires $d-1$ new indices to become $R_{1,d+1,\cdots, 2d-1}$. The third operator begins with the index 2 and acquires $d-2$ new indices to become $R_{2,d+1,2d,\cdots, 3d-3}$. This pattern can be extended to find that the $k$th operator begins with the index $k-1$ and acquires $d-k+1$ new indices to become $R_{k-1,d+k-2,\cdots, kd-\frac{k(k-1)}{2}}$. The values of $k$ run up to $d+1$. The right hand side of the equation is just the left hand side in reverse order.

The $d$-simplex equations written here do not depend on spectral parameters, or are the constant versions. Like in the Yang-Baxter equation it is physically important for the $d$-simplex equations to depend on spectral parameters that appear in the scattering processes. This dependence will be shown in the next section when we construct their solutions. 

\section{Solutions}
\label{sec:solutions}
We will now present an algebraic method to generate solutions of the higher simplex equations. For each $d$ we solve the vertex form of the higher simplex equation as described in \cite{MAILLET1989221}. We call the solutions as $d$-simplex operators. The technique is a generalisation of the one used in \cite{padmanabhan2024integrability}. As a result we will see that the latter reduce to special cases of the solutions presented here. 

The ansatzes for the $d$-simplex operators are constructed with two operators $A$ and $B$. These operators act on a space $V$, which is the local Hilbert space labeled by the indices appearing in the higher simplex equations. We assume they satisfy 
\begin{equation}\label{eq:ABalpha}
    AB = \alpha BA,
\end{equation}
for some parameter $\alpha\in\mathbb{C}$. Additionally the operators $A$ and $B$ can depend on a set of parameters. For simplicity we will ignore them in the proofs to follow. These ansatzes can also be generalised to include more number of local operators as we shall see.

The technique is well illustrated for the 2-simplex equation or the {\bf Yang-Baxter equation}. This will serve as a warm-up for the higher simplex cases. We begin with two simple solutions of the Yang Baxter equation\footnote{Such solutions are discussed in \cite{padmanabhan2024integrability}.} :
$$ A_iA_j~;~B_iB_j.$$
Next we consider the $R$-matrix, 
\begin{equation}\label{eq:Rd2}
    R_{ij} = A_iA_j + B_iB_j. 
\end{equation}
This ansatz is plugged into the constant [independent of spectral parameter]\footnote{This is also known as the braid relation.} Yang-Baxter equation,
\begin{equation}\label{eq:d2simplex}
    R_{12}R_{13}R_{23} = R_{23}R_{13}R_{12}.
\end{equation}
This is the vertex form of the Yang-Baxter equation. Now we check the conditions on $\alpha$ for \eqref{eq:Rd2} to satisfy \eqref{eq:d2simplex}.
\begin{eqnarray}
    R_{12}R_{13}R_{23} & = & A_2A_3\left(A_1A_2 + \frac{1}{\alpha}B_1B_2 \right)\left(A_1A_3 + \frac{1}{\alpha}B_1B_3\right) \nonumber \\
    & + & B_2B_3\left(\alpha A_1A_2 + B_1B_2\right)\left(\alpha A_1A_3 + B_1B_3 \right) \nonumber \\
    & = & \left(A_2A_3 + \alpha^2 B_2B_3\right)\left(A_1A_2 + \frac{1}{\alpha}B_1B_2 \right)\left(A_1A_3 + \frac{1}{\alpha}B_1B_3\right) \nonumber \\
    & = & \left(A_2A_3 + \alpha^2 B_2B_3\right)\left[A_1A_3\left(A_1A_2 + \frac{1}{\alpha^2}B_1B_2\right) \right. \nonumber \\
    & + & \left. B_1B_3\left(A_1A_2 + \frac{1}{\alpha^2}B_1B_2 \right) \right] \nonumber \\
    & = & \left(A_2A_3 + \alpha^2 B_2B_3\right)R_{13}\left(A_1A_2 + \frac{1}{\alpha^2}B_1B_2 \right) \nonumber \\
    & = & R_{23}R_{13}R_{12}~;~\textrm{iff}~\alpha^2=1.
\end{eqnarray}
Thus we find that when  
\begin{equation}
    \alpha=\pm 1
\end{equation}
the $R$-matrix in \eqref{eq:Rd2} satisfies the constant Yang-Baxter equation \eqref{eq:d2simplex}. We note that the $\alpha=1$ case, corresponding to commuting $A$ and $B$, are the solutions considered in \cite{padmanabhan2024integrability}. When $\alpha=-1$, the operators anticommute, which can be realised by Clifford algebras of arbitrary orders. Note that the set of solutions found above form a {\bf linear space}. Thus linear combinations of the two operators in \eqref{eq:Rd2} 
\begin{equation}\label{eq:Rd2nonadditive}
    R_{ij}(\mu_i, \mu_j) = \mu_i A_iA_j + \mu_j B_iB_j,
\end{equation}
satisfy the Yang-Baxter equation. 
\begin{equation}\label{eq:d2simplexSP}
    R_{12}(\mu_1, \mu_2)R_{13}(\mu_1, \mu_3)R_{23}(\mu_2, \mu_3) = R_{23}(\mu_2, \mu_3)R_{13}(\mu_1, \mu_3)R_{12}(\mu_1, \mu_2).
\end{equation}
This is the Yang-Baxter equation in the non-additive form. Here $\mu$'s are some complex numbers, we can identify them  with the spectral parameters. Note that linear combinations of the form
\begin{equation}\label{eq:Rd2gen-constant}
    R_{ij} = \mu A_iA_j + \nu B_iB_j
\end{equation}
with the coefficients not depending on the site indices, satisfy the constant Yang-Baxter equation \eqref{eq:d2simplex}.


We can further generalise these solutions by including more local operators. Consider two sets of mutually anticommuting operators :
\begin{equation}\label{eq:ABClifford}
    \{A^{(m)}\vert m\in \left(1,\cdots, r\right) \}~;~\{B^{(n)}\vert n\in \left(1,\cdots, s\right) \}.
\end{equation}
These operators satisfy,
\begin{eqnarray}
    \left[A^{(m_1)}, A^{(m_2)}\right] = \left[B^{(n_1)}, B^{(n_2)}\right] = \left\{A^{(m_1)}, B^{(n_1)}\right\} = 0~;~ \forall~m_1, m_2, n_1, n_2.
\end{eqnarray}
Using the linear structure of the solutions we have the following $R$-matrix
\begin{equation}\label{eq:Rd2general}
    R_{ij} = \sum\limits_{m_1, m_2=1}^r~\mu_{m_1m_2}\left(A^{(m_1)}\right)_i\left(A^{(m_2)}\right)_j + \sum\limits_{n_1, n_2=1}^s~\nu_{n_1n_2}\left(B^{(n_1)}\right)_i\left(B^{(n_2)}\right)_j,
\end{equation}
as the most general Clifford solution to the 2-simplex equation in the non-additive form. This solutions contains $r^2+s^2$ parameters.

Next we consider operators of the form :
$$ A_iB_j~;~B_iA_j.$$
Assuming that $A$ and $B$ anticommute each of these solve what we call the anti-Yang-Baxter equation as described in Appendix \ref{app:antidsimplex}. We also see that their linear combinations satisfy the same. As with the Yang-Baxter equation we can solve both the constant and spectral parameter dependent anti-Yang-Baxter equation using these operators.

{\it {\bf Summary :}} We will now organise the 2-simplex (anti-2-simplex) or Yang-Baxter (anti-Yang-Baxter) operators. They are made out of factorised operators consisting of the anticommuting $A$ and $B$ operators on the two sites. We have three types of such operators classified by the number of $A$'s and $B$'s occurring in them\footnote{We use this description for the higher simplex cases as well.}. Denoting these numbers as $a$ and $b$ the classification is summarised in Table \ref{tab:2simplexOP}.
\begin{table}[h!]
\centering
\begin{tabular}{ |c|c| } 
 \hline
 Type & Operators  \\
 \hline
 \hline
 $(2,0)$ & $A_iA_j$  \\
 \hline
 $(1,1)$ & $A_iB_j$, $B_iA_j$ \\ 
 \hline
 $(0,2)$ & $B_iB_j$ \\
 \hline
 \end{tabular}
 \caption{The different building blocks of the Yang-Baxter (anti-Yang-Baxter) operators.}
 \label{tab:2simplexOP}
\end{table}
The Yang-Baxter operators are made from linear combinations of the $(2,0)$ and $(0,2)$ type operators whereas linear combinations of the $(1,1)$ type operators solve the anti-Yang-Baxter equation (See Appendix \ref{app:antidsimplex}). Combining all three types produces operators that solve a ``Yang-Baxter-like'' equation. We will not study this in this paper. This completes our analysis of the $d=2$ or the Yang-Baxter case.

Henceforth we will assume that the local operators $A$ and $B$ anticommute while considering higher simplex operators. Next we show that such anticommuting operators can naturally be constructed out of basis elements of a Clifford algebra. The method to generate the $d$-simplex operators only assumes that $A$ and $B$ anticommutes and thus the following subsection can be skipped by the reader who wishes to go directly to the solutions.

\subsection{Clifford algebras}
\label{subsec:clifford}
The operators $A$ and $B$ appearing in \eqref{eq:ABalpha} are required to either commute or anticommute for the $R$-matrix in \eqref{eq:Rd2} to satisfy the 2-simplex or the constant Yang-Baxter equation. The commuting case is the subject of \cite{padmanabhan2024integrability}. The commuting case also satisfies the higher simplex equations. The anticommuting case will be the focus of this paper. The condition of anticommutativity places constraints on the operators $A$ and $B$. There are three possibilities.
\begin{enumerate}
    \item  Both $A$ and $B$ are invertible.
    \item  One of them is invertible and the other non-invertible.
    \item Both $A$ and $B$ are non-invertible.
\end{enumerate}
\paragraph{\textbf{Case 1} :} We assume $A$ and $B$ are invertible and they generate a closed algebra. The algebra is taken to be finite dimensional. Then we have 
\begin{equation}\label{eq:AinvB}
    AB = -BA \implies A^{-1}B = -BA^{-1}.
\end{equation}

As $A$ and $B$ form a closed algebra, the operator $A^{-1}$ has to be a linear combination of $A$ and $B$\footnote{A more general assumption here can be $A^{-1}=\alpha~A+\beta~B+\gamma~AB$. Using \eqref{eq:AinvB} and $AA^{-1}=A^{-1}A=\mathbb{1}$, we can show that both $\beta=\gamma=0$.}, 
\begin{equation}
    A^{-1} = \alpha~A + \beta~B.
\end{equation}
Substituting this into \eqref{eq:AinvB} we find that 
\begin{equation}
    \beta B^2 = 0 \implies \beta = 0.
\end{equation}
Thus $A^{-1}\propto A$. We take the proportionality constant to be $\alpha=\pm 1$ without loss of generality. Any other positive or negative constant can be absorbed into $A$, thus scaling $\alpha$ to $\pm 1$. The above arguments imply 
\begin{equation}
    A^2 = \pm \mathbb{1}.
\end{equation}
The roles of $A$ and $B$ can be reversed in \eqref{eq:AinvB} to show that 
\begin{equation}
     B^2 = \pm \mathbb{1}.
\end{equation}
Thus we find that when two invertible operators $A$ and $B$, forming a closed algebra, anticommute with each other, their squares are either $\pm \mathbb{1}$. Operators satisfying these relations form a Clifford algebra.
\begin{definition}
    Consider a vector space $V$ with a degenerate quadratic form denoted by a multiplication. The quadratic form has a signature $p+q$ with $p$ and $q$ being positive integers. A Clifford algebra is an associative algebra generated by $p+q$ elements denoted $$\{\Gamma_1, \cdots , \Gamma_{p+q}\}.$$ They satisfy the relations
    \begin{eqnarray}
        \Gamma^2_i & = & \mathbb{1}~\textrm{for}~1\leq i\leq p, \label{eq:clifford-1} \\ 
        \Gamma^2_i & = & -\mathbb{1}~\textrm{for}~p+1\leq i\leq p+q,\label{eq:clifford-2}\\ 
        \Gamma_i\Gamma_j & = & -\Gamma_j\Gamma_i~\textrm{for}~i\neq j. \label{eq:clifford-3}
    \end{eqnarray}
 We denote the Clifford algebra by $\mathbf{CL}(p,q)$ with order $p+q$. 
\end{definition}
The identity operator $\mathbb{1}$ is also included in the algebra. It is obtained from the generators. The vector space can be real or complex. This matters for the representations of $\mathbf{CL}(p,q)$.

The Clifford relations \eqref{eq:clifford-1}-\eqref{eq:clifford-3} help determine the dimension of $\mathbf{CL}(p,q)$. An arbitrary element of the basis element is an unordered word in the generators. This can be written as $\Gamma_1^{j_1}\cdots\Gamma_{p+q}^{j_{p+q}}$ with $\left(j_1,\cdots, j_{p+q}\right)\in\{0,1 \}.$ Thus each generator is included or not in each word making the total number of words $2^{p+q}=\textrm{dim}~\mathbf{CL}(p,q)$. Each word (element) of the Clifford algebra has a definite grade.
\begin{definition}
    The grade of an element or word of $\mathbf{CL}(p,q)$ is the number of unique generators needed to generate the word.
\end{definition}
For example the scalars are grade 0, the set of generators ($\Gamma_i$) are of grade 1. Bilinears\footnote{For real Clifford algebras these are also called bivectors and have a natural interpretation in geometric algebra.} in the generators ($\Gamma_i\Gamma_j$) are of grade 2 and so on.
We can now write down the basis elements of $\mathbf{CL}(p,q)$ by distinguishing words according to their grade. The different spaces along with their dimensions are
\begin{eqnarray}
    G_0 & = & k~\mathbb{1}~k\in\mathbb{C}~\textrm{or}~\mathbb{R},~\textrm{dim}=\binom{p+q}{0} \nonumber \\
    G_1 & = & \{\Gamma_i\vert i\in\{1,\cdots, p+q\} \},~\textrm{dim}=\binom{p+q}{1} \nonumber \\
    G_2 & = & \{\Gamma_i\Gamma_j\vert i\neq j\in\{1,\cdots, p+q\} \},~\textrm{dim}=\binom{p+q}{2} \nonumber \\
    & \vdots & \nonumber \\
    G_{p+q} & = & \Gamma_1\cdots\Gamma_{p+q},~\textrm{dim}=\binom{p+q}{p+q}.
\end{eqnarray}
The sum of the dimensions of each of these graded spaces
\begin{equation}
    \sum\limits_{j=0}^{p+q}~\binom{p+q}{j} = 2^{p+q}
\end{equation}
as expected. With this the Clifford algebra becomes
\begin{equation}
    \mathbf{CL}(p,q) = G_0 \oplus G_1 \oplus G_2 \oplus \cdots \oplus G_{p+q}.
\end{equation}
The different elements of $\mathbf{CL}(p,q)$ are either even or odd. In particular we find that the even elements of the Clifford algebra form a subalgebra. The grading structure is useful to identify the mutually anticommuting operators in \eqref{eq:ABClifford} with elements of $\mathbf{CL}(p,q)$. Let us consider two examples. 

{\it Example 1 :} Consider a Clifford algebra of order 2. This can be one of three possibilities - $\mathbf{CL}(2,0)$, $\mathbf{CL}(1,1)$, $\mathbf{CL}(0,2)$. Barring the identity operator there are three non-trivial elements $\{\Gamma_1, \Gamma_2, \Gamma_1\Gamma_2\}$. Any of the three Clifford algebras will generate $R$-matrices in \eqref{eq:Rd2}. This is done by choosing the $A$ and $B$ operators as shown in Table \ref{tab:ABcl2}.
\begin{table}[h!]
\centering
\begin{tabular}{ |c|c| } 
 \hline
 $A$ & $B$  \\
 \hline
 \hline
 $\Gamma_1$ & $\Gamma_2$  \\
 \hline
 $\Gamma_1$ & $\Gamma_1\Gamma_2$ \\ 
 \hline
 $\Gamma_2$ & $\Gamma_1\Gamma_2$ \\
 \hline
 \end{tabular}
 \caption{Possible choices of the $A$ and $B$ operators from a Clifford algebra of order 2.}
 \label{tab:ABcl2}
\end{table}
When using Clifford algebras of order 2 there are not enough operators to construct the $R$-matrices in \eqref{eq:Rd2general}. For this we need to use Clifford algebras of order higher than 2. 

{\it Example 2 :} For this case we choose a Clifford algebra of order 3. There are four such examples - $\mathbf{CL}(3,0)$, $\mathbf{CL}(2,1)$, $\mathbf{CL}(1,2)$ and $\mathbf{CL}(0,3)$. The basis elements are given by
$$\{\mathbb{1}~;~\Gamma_1, \Gamma_2, \Gamma_3~;~\Gamma_1\Gamma_2, \Gamma_1\Gamma_3, \Gamma_2\Gamma_3~;~\Gamma_1\Gamma_2\Gamma_3 \}. $$

The choices of $A$ and $B$ resulting in the $R$-matrix in \eqref{eq:Rd2} are shown in Table \ref{tab:ABcl3-1}.
\begin{table}[h!]
\centering
\begin{tabular}{ |c|c| } 
 \hline
 $A$ & $B$  \\
 \hline
 \hline
 $\Gamma_1$ & $\Gamma_2$/$\Gamma_3$/$\Gamma_1\Gamma_2$/$\Gamma_1\Gamma_3$  \\
 \hline
 $\Gamma_2$ & $\Gamma_3$/$\Gamma_1\Gamma_2$/$\Gamma_2\Gamma_3$ \\ 
 \hline
 $\Gamma_3$ & $\Gamma_1\Gamma_3$/$\Gamma_2\Gamma_3$ \\
 \hline
 \end{tabular}
 \caption{Possible choices of the $A$ and $B$ operators for the $R$-matrix in \eqref{eq:Rd2} from a Clifford algebra of order 3. Interchanging $A$ and $B$ gives another set of $R$-matrices as discussed in the text.}
 \label{tab:ABcl3-1}
\end{table}
Next we realise the anticommuting sets of operators in \eqref{eq:ABClifford} using the basis elements of an order 3 Clifford algebra. This helps us construct the $R$-matrices in \eqref{eq:Rd2general}. Note that we cannot use the grade 3 element $\Gamma_1\Gamma_2\Gamma_3$ in any of the sets as this element is central\footnote{An analogous statement holds for all Clifford algebras of odd order.}. The choice of the $A$'s and $B$'s for $r=s=2$ is shown in Table \ref{tab:ABcl3-2}.
\begin{table}[h!]
\centering
\begin{tabular}{ |c|c| } 
 \hline
 $A$ & $B$  \\
 \hline
 \hline
 $\left(\Gamma_1, \Gamma_2\Gamma_3\right)$ & $\left(\Gamma_3, \Gamma_1\Gamma_2\right)$  \\
 \hline
 $\left(\Gamma_1, \Gamma_2\Gamma_3\right)$ & $\left(\Gamma_2, \Gamma_1\Gamma_3\right)$ \\ 
 \hline
 $\left(\Gamma_2, \Gamma_1\Gamma_3\right)$ & $\left(\Gamma_3, \Gamma_1\Gamma_2\right)$ \\
 \hline
 \end{tabular}
 \caption{Possible choices of the $A$'s and $B$'s for $r=s=2$ to construct the $R$-matrix in \eqref{eq:Rd2general} from a Clifford algebra of order 3. Interchanging $A$ and $B$ gives another set of $R$-matrices as discussed in the text.}
 \label{tab:ABcl3-2}
\end{table}

\paragraph{\textbf{Case 2} :} Let $A$ be non-invertible and $B$ invertible\footnote{Interchanging the roles of $A$ and $B$ does not produce anything new.}. From the arguments in case 1 we can show that
$$ B^2 = \pm \mathbb{1}.$$
As $A$ and $B$ anticommute we have 
\begin{equation}\label{eq:A2B}
    A^2B = BA^2.
\end{equation}
Since $A$ is non-invertible, its determinant is 0. We consider two possibilities\footnote{We are being conservative here as there can be more possibilities for a non-invertible matrix. However for 2 by 2 matrices, which we will mostly interested in, these are the only possibilities.} : 
\begin{enumerate}
    \item $A$ is nilpotent : $A^2=0$.
    \item $A$ is a projector : $A^2=A$.
\end{enumerate}
For the first case we consider Clifford algebras generalised to include nilpotent operators. The new definition involves a set of $y$ generators that are nilpotent. The resulting Clifford algebra will be denoted $\mathbf{CL}(p,q,y)$\footnote{The set of nilpotent anticommuting elements is also called the Grassmann algebra.}. Thus they can be used to realise the $A$ and $B$ operators. 

We could also realise the first case with Clifford algebras of the type $\mathbf{CL}(p,q)$. As an example consider an order 2 Clifford algebra, $\mathbf{CL}(2,0)$. Then
\begin{equation}
    A = \Gamma_1 + \Gamma_1\Gamma_2~;~B=\Gamma_2,
\end{equation}
satisfies the conditions required of $A$ and $B$.

In the second case \eqref{eq:A2B} reduces to
\begin{equation}
    AB  = BA =0.
\end{equation}
But since $B$ is invertible the only way this equation can be satisfied is when $A=0$. This is a trivial solution.

Finally we consider another situation. A non-invertible operator $A$ should satisfy a characteristic equation of the form
$$A\left(\sum\limits_n~k_nA^n\right) = 0. $$
From this substitute for $A^2$ into \eqref{eq:A2B}. We find that the even coefficients of this sum are 0. We are then left with two cases.  
\begin{enumerate}
    \item $A$ is nilpotent : $A^2=0$.
    \item $A$ satisfies $\sum\limits_{n\in\textrm{odd}}~k_nA^{n-1}=0$.
\end{enumerate}
The first case has already been considered. The characteristic equation in the second case can be chosen in a particular representation. In this case the $A$ operators can still be expanded in terms of Clifford basis elements.

\paragraph{\textbf{Case 3} :} Next we turn to the case when $A$ and $B$ are not invertible. This implies that \eqref{eq:AinvB} no longer holds. However \eqref{eq:A2B} is still true. From the arguments of Case 2, we have that 
\begin{equation}
    A^2B = 0 ~;~AB^2 =0.
\end{equation}
Subtracting these two we obtain 
\begin{equation}
    \left(A+B\right)AB=0.
\end{equation}
We want $A$ and $B$ to be independent of each other. So they must satisfy $AB=0$. The operators $A$ and $B$ can either be projectors or they can be nilpotent. When the operators are projectors they cannot be thought of as generators of a Clifford algebra. Nevertheless we can realise such operators using the basis elements of a Clifford algebra of an appropriate order. 
As an example we realise the two operators from the order 2 Clifford algebra, $\mathbf{CL}(2,0)$ to check these situations.

{\it Example 1 :} Consider $A$ as nilpotent and $B$ as a projector. We require $AB=BA=0$. This is not possible as $A$ and $B$ will have to be orthogonal to each other, but we have taken one of them to be nilpotent. So this is a contradiction and hence this case is ruled out.

{\it Example 2 :} Both $A$ and $B$ are projectors. Their product is zero when they are orthogonal to each other,
\begin{equation}
    A =\frac{\mathbb{1}+\Gamma_1}{2} ~;~B = \frac{\mathbb{1}-\Gamma_1}{2}.
\end{equation}

{\it Example 3 :} The third situation where $A$ and $B$ are realised using $\mathbf{CL}(2,0)$ and are both nilpotent is not possible. However if we use $\mathbf{CL}(3,0)$ we can find such $A$'s and $B$'s. One example is 
\begin{equation}
    A = \Gamma_1-\mathrm{i}\Gamma_3~;~B = \Gamma_1\Gamma_2 + \mathrm{i}\Gamma_2\Gamma_3,
\end{equation}
with $\mathrm{i}=\sqrt{-1}$.

\subsubsection{Summary of the Clifford algebra}
\label{subsubsec:summary}
We considered the situations when two operators $A$ and $B$ anticommute. There are three possibilities. 
\begin{enumerate}
    \item {\it Case 1 :} When $A$ and $B$ are both invertible, $A^2=\pm\mathbb{1}$ and $B^2=\pm\mathbb{1}$. Then both $A$ and $B$ satisfy the relations of a Clifford algebra $\mathbf{CL}(p,q)$. 
    \item {\it Case 2 :} When $A$ is non-invertible and $B$ is invertible, $A^2=0$ and $B^2=\pm\mathbb{1}$. The operators can be realised using a more general Clifford algebra that also includes nilpotent generators, $\mathbf{CL}(p,q,r)$. The nilpotent operator $A$ can also be obtained using the basis elements of $\mathbf{CL}(p,q)$.
    \item {\it Case 3 :} When both $A$ and $B$ are non-invertible we have $AB=0$. This is non-trivially satisfied when they are either nilpotent or they are projectors. In all the cases the Clifford algebra $\mathbf{CL}(p,q)$ can be used to realise them.
\end{enumerate}

\subsection{$3$ - Simplex or Zamalodchikov's tetrahedron equations}
\label{subsec:d3}
The 3-simplex equation or the tetrahedron equation is the first higher simplex equation that will admit solutions constructed out of Clifford algebras\footnote{As with the $d=2$ or Yang-Baxter equation we could consider an ansatz where the $\alpha$ in \eqref{eq:ABalpha} is not $\pm 1$. It turns out that the 3-simplex equation is satisfied only when $\alpha=\pm 1$. This is shown in Appendix \ref{app:ABalphad3}. }. As in the 2-simplex case we will demonstrate this with some simple ansatzes. Subsequently we will generalise this by identifying the conditions for these solutions to form a linear space.

Consider the 3-simplex operator,
\begin{equation}\label{eq:Rd3product}
    R_{ijk} = A_iA_jB_k.
\end{equation}
This solves the constant 3-simplex equation in the vertex form,
\begin{equation}\label{eq:d3simplex}
    R_{123}R_{145}R_{246}R_{356} = R_{356}R_{246}R_{145}R_{123},
\end{equation}
as both sides of this equation simplify to 
$$A_1^2A_2^2\left(B_3A_3\right)A_4^2\left(B_5A_5\right)A_6^2.$$
In a similar vein the choices
$$ A_iB_jA_k~,~B_iA_jA_k,$$
also satisfy the tetrahedron equation in \eqref{eq:d3simplex}. Notice that the three operators $A_iA_jB_k$, $A_iB_jA_k$ and $B_iA_jA_k$ commute with each other. Each of them contains two $A$'s and one $B$ operator and so they are of the $(2,1)$ type. (See Table \ref{tab:2simplexOP} for analogous notation in the $d=2$ case). Naively we expect a linear combination of these three operators to also be a solution to the tetrahedron equation \eqref{eq:d3simplex}. We check this systematically starting with the sum
\begin{equation}\label{eq:Rd3sum}
   R_{ijk} =  A_iA_jB_k + A_iB_jA_k.
\end{equation}
The proof that this solves the tetrahedron equation \eqref{eq:d3simplex} goes as follows :
\begin{eqnarray}
   & & R_{123}R_{145}R_{246}R_{356} \nonumber \\
   & = & A_3A_5B_6\left(-A_1A_2B_3 + A_1B_2A_3\right)\left(-A_1A_4B_5 + A_1B_4A_5\right)\left(A_2A_4B_6 - A_2B_4A_6\right) \nonumber \\
    & + & A_3B_5A_6\left(-A_1A_2B_3 + A_1B_2A_3\right)\left(A_1A_4B_5 - A_1B_4A_5\right)\left(-A_2A_4B_6 + A_2B_4A_6\right) \nonumber \\
    & = & R_{356}\left(-A_1A_2B_3 + A_1B_2A_3\right)\left(-A_1A_4B_5 + A_1B_4A_5\right)\left(A_2A_4B_6 - A_2B_4A_6\right) \nonumber \\
    & = & R_{3456}\left[A_2A_4B_6\left(-A_1A_2B_3 - A_1B_2A_3\right)\left(-A_1A_4B_5 - A_1B_4A_5\right) \right. \nonumber \\
    &-& \left.A_2B_4A_6\left(-A_1A_2B_3 - A_1B_2A_3\right)\left(A_1A_4B_5 + A_1B_4A_5\right) \right] \nonumber \\
    & = & R_{356}R_{246}\left(A_1A_2B_3 + A_1B_2A_3\right)\left(A_1A_4B_5 + A_1B_4A_5\right) \nonumber \\
    & = & R_{356}R_{246}R_{145}R_{123}.
\end{eqnarray}
This confirms the linear structure. The nature of the proof shows that the linear structure will hold for the entire set of three product 3-simplex operators. We check this for the next generalisation,
\begin{equation}\label{eq:Rd3}
    R_{ijk} = A_iA_jB_k + A_iB_jA_k + B_iA_jA_k.
\end{equation}
The proof that this solves the tetrahedron equation is straightforward but the notation makes it cumbersome. To simplify things we introduce a new notation.

\subsubsection*{Notation to simplify proofs}
\label{subsubsec:notation}
The proofs in the rest of the paper are virtually impossible to write down without some form of shorthand notation. We will introduce this for the 3-simplex operators. The generalisation to the higher simplex cases is straightforward.
\begin{equation}\label{eq:Rd3shorthand}
    R_{ijk}(\kappa_1,\kappa_2,\kappa_3) = \kappa_1~A_iA_jB_k + \kappa_2~A_iB_jA_k + \kappa_3~B_iA_jA_k\equiv \left( \kappa_1,\kappa_2,\kappa_3\right)_{ijk}~;~\kappa_1,\kappa_2,\kappa_3\in\{+,-\}.
\end{equation}
The $\kappa_j$'s are the coefficients of the terms appearing in the 3-simplex operator. For the ansatz in \eqref{eq:Rd3} the coefficients are just one and so in this case the $\kappa_j$'s specify their signs\footnote{In all of the proofs to follow only this sign, and not the coefficient of these operators, matters.}.
Thus the shorthand for this 3-simplex operator becomes  $\left(+,+,+\right)_{ijk}$. Note that the $\kappa_j$'s are not to be confused with spectral parameters. The sole purpose of their introduction is to simplify the upcoming proofs. We are now ready to check that \eqref{eq:Rd3} satisfies the 3-simplex equation \eqref{eq:d3simplex} :
\begin{eqnarray}
   R_{123}R_{145}R_{246}R_{356}  & = & A_3A_5B_6\left(-,+,+\right)_{123}\left(-,+,+\right)_{145}\left(+,-,- \right)_{246} \nonumber \\
    & + & A_3B_5A_6\left(-,+,+\right)_{123}\left(+,-,-\right)_{145}\left(-,+,+\right)_{246} \nonumber \\
    & + & B_3A_5A_6\left(+,-,-\right)_{123}\left(-,+,+\right)_{145}\left(-,+,+\right)_{246} \nonumber \\
    & = & R_{356}\left(+,-,-\right)_{123}\left(+,-,-\right)_{145}\left(+,-,-\right)_{246} \nonumber \\
    & = & R_{356}\left[A_2A_4B_6\left(+,+,-\right)_{123}\left(+,+,- \right)_{145} \right. \nonumber \\
    & - & \left.A_2B_4A_6\left(+,+,-\right)_{123}\left(-,-,+\right)_{145} \right.\nonumber \\
    & - & \left.B_2A_4A_6\left(-,-,+\right)_{123}\left(+,+,-\right)_{145} \right]\nonumber \\
    & = & R_{356}R_{246}\left(+,+,-\right)_{123}\left(+,+,-\right)_{145} \nonumber \\
    & = & R_{356}R_{246}\left[A_1A_4B_5\left(+,+,+\right)_{123} \right. \nonumber \\
    & + & \left. A_1B_4A_5\left(+,+,+\right)_{123} \right. \nonumber \\
    & - & \left. B_1A_4A_5\left(-,-,-\right)_{123} \right] \nonumber \\
    & = & R_{356}R_{246}R_{145}R_{123}.
\end{eqnarray}
This establishes the linearity of these solutions. Thus we can generalise the 3-simplex operator \eqref{eq:Rd3} by taking linear combinations of the three operators in the sum in \eqref{eq:Rd3} to get 
\begin{equation}\label{eq:Rd3nonadditive}
    R_{ijk}(\mu_i, \mu_j, \mu_k) = \mu_i~A_iA_jB_k + \mu_j~A_iB_jA_k + \mu_k~B_iA_jA_k,
\end{equation}
where the $\mu$'s are identified as spectral parameters.
This satisfies a spectral parameter dependent 3-simplex equation
\begin{eqnarray}\label{eq:d3simplexSP}
  & &  R_{123}\left(\mu_1, \mu_2, \mu_3\right)R_{145}\left(\mu_1, \mu_4, \mu_5\right)R_{246}\left(\mu_2, \mu_4, \mu_6\right)R_{356}\left(\mu_3, \mu_5, \mu_6\right) \nonumber \\
  & = & R_{356}\left(\mu_3, \mu_5, \mu_6\right)R_{246}\left(\mu_2, \mu_4, \mu_6\right)R_{145}\left(\mu_1, \mu_4, \mu_5\right)R_{123}\left(\mu_1, \mu_2, \mu_3\right).
\end{eqnarray}
When the coefficients $\mu$ are not site dependent this operator solves the constant tetrahedron equation \eqref{eq:d3simplex}.
Next we take an arbitrary linear combination of the two sets of anticommuting operators in \eqref{eq:ABClifford} to obtain the most general 3-simplex operator constructed out of Clifford algebras
\begin{eqnarray}\label{eq:Rd3general}
   R_{ijk}\left(\bm{\mu}_i, \bm{\nu}_j, \bm{\omega}_k\right) & = &  \sum\limits_{m_1,m_2=1}^r\sum\limits_{n=1}^s~\left[\left(\mu_{m_1m_2n}\right)_i\left(A^{(m_1)}\right)_i\left(A^{(m_2)}\right)_j\left(B^{(n)}\right)_k \right. \nonumber \\ & + & \left. \left(\nu_{m_1m_2n}\right)_j\left(A^{(m_1)}\right)_i\left(B^{(n)}\right)_j\left(A^{(m_2)}\right)_k \right. \nonumber \\ & + & \left. \left(\omega_{m_1m_2n}\right)_k\left(B^{(n)}\right)_i\left(A^{(m_1)}\right)_j\left(A^{(m_2)}\right)_k\right]. 
\end{eqnarray}
Here the spectral parameter $\bm{\mu}$ is the tuple $\mu_{m_1m_2n}$ with $r^2s$ elements. Thus the solution in \eqref{eq:Rd3general} contains a total of $3r^2s$ parameters. It satisfies a more general form of the spectral parameter dependent 3-simplex equation
\begin{eqnarray}\label{eq:d3simplexSPgeneral}
  & &  R_{123}\left(\bm{\mu}_1, \bm{\nu}_2, \bm{\omega}_3\right)R_{145}\left(\bm{\mu}_1, \bm{\nu}_4, \bm{\omega}_5\right)R_{246}\left(\bm{\mu}_2, \bm{\nu}_4, \bm{\omega}_6\right)R_{356}\left(\bm{\mu}_3, \bm{\nu}_5, \bm{\omega}_6\right) \nonumber \\
  & = & R_{356}\left(\bm{\mu}_3, \bm{\nu}_5, \bm{\omega}_6\right)R_{246}\left(\bm{\mu}_2, \bm{\nu}_4, \bm{\omega}_6\right)R_{145}\left(\bm{\mu}_1, \bm{\nu}_4, \bm{\omega}_5\right)R_{123}\left(\bm{\mu}_1, \bm{\nu}_2, \bm{\omega}_3\right).
\end{eqnarray}
Two remarks are in order. 

{\it {\bf Remark 1 :}} Interchanging the $A$ and $B$ operators in \eqref{eq:Rd3} leads to another solution,
\begin{eqnarray}\label{eq:Rd3generalBA}
   \tilde{R}_{ijk}\left(\bm{\mu}_i, \bm{\nu}_j, \bm{\omega}_k\right) & = &  \sum\limits_{n_1,n_2=1}^s\sum\limits_{m=1}^r~\left[\left(\mu_{n_1n_2m}\right)_i\left(B^{(n_1)}\right)_i\left(B^{(n_2)}\right)_j\left(A^{(m)}\right)_k \right. \nonumber \\ & + & \left. \left(\nu_{n_1n_2m}\right)_j\left(B^{(n_1)}\right)_i\left(A^{(m)}\right)_j\left(B^{(n_2)}\right)_k \right. \nonumber \\ & + & \left. \left(\omega_{n_1n_2m}\right)_k\left(A^{(m)}\right)_i\left(B^{(n_1)}\right)_j\left(B^{(n_2)}\right)_k\right]. 
\end{eqnarray}
This operator is made of the $(1,2)$ type operators, $$ B_iB_jA_k~;~B_iA_jB_k~;~A_iB_jB_k,$$ that mutually commute with each other.
This ansatz satisfies the spectral parameter dependent 3-simplex equation in \eqref{eq:d3simplexSPgeneral}. When $r=s=1$, this solution can be obtained by interchanging $A$ and $B$ in \eqref{eq:Rd3nonadditive} by an invertible operator. When this is the case the two 3-simplex operators are equivalent to each other by a local invertible operator\footnote{These two $R$-matrices are gauge equivalent in old terminology.} :
\begin{equation}
    \tilde{R}_{ijk}(\mu_i,\mu_j,\mu_k) = \kappa~Q_iQ_jQ_k~R_{ijk}(\mu_i,\mu_j,\mu_k)~Q_k^{-1}Q_j^{-1}Q_i^{-1},
\end{equation}
where $\kappa$ is a complex number. It is easy to see that $\tilde{R}$ also satisfies the 3-simplex equation \eqref{eq:d3simplexSP}. Due to this equivalence we do not consider $\tilde{R}$ and $R$ to be different solutions of the 3-simplex equation.  
When no such $Q$ exists the two 3-simplex operators are taken to be inequivalent solutions.
We will see explicit examples of both these situations in Sec. \ref{sec:qubitsolutions}. The above arguments can be extended for the case when $r=s\neq 1$. For example consider the case $r=s=2$. Possible choices for $A$ and $B$ are listed in Table \ref{tab:ABcl3-2}. The operator $\frac{\Gamma_1+\Gamma_3}{\sqrt{2}}$ interchanges the two columns in the first row of Table \ref{tab:ABcl3-2}. This is the analog of the Hadamard gate for an order 3 Clifford algebra.  

{\it {\bf Remark 2 :}}
An alternate choice to the ansatz in \eqref{eq:Rd3product} is
\begin{equation}\label{eq:Rd3product-anti}
   R_{ijk}=A_iA_jA_k. 
\end{equation}
This is easily seen to solve the tetrahedron equation in \eqref{eq:d3simplex}. In a similar manner the operator $B_iB_jB_k$ also solves the tetrahedron equation. These operators are of the type $(3,0)$ and $(0,3)$ respectively. Such a choice exists for all the $d$-simplex operators. These choices anticommute with each other. Hence we do not expect a linear combination of the type $$ A_iA_jA_k + B_iB_jB_k$$ to solve the tetrahedron equation of \eqref{eq:d3simplex}. Nevertheless it solves the anti-tetrahedron equation as discussed in Appendix \ref{app:antidsimplex}.

{\it {\bf Summary :}} We summarise the 3-simplex or tetrahedron operators in Table \ref{tab:3simplexOP}.
\begin{table}[h!]
\centering
\begin{tabular}{ |c|c| } 
 \hline
 Type & Operators  \\
 \hline
 \hline
 $(3,0)$ & $A_iA_jA_k$  \\
 \hline
 $(2,1)$ & $A_iA_jB_k$, $A_iB_jA_k$, $B_iA_jA_k$ \\ 
 \hline
 $(1,2)$ & $B_iB_jA_k$, $B_iA_jB_k$, $A_iB_jB_k$ \\
 \hline
 $(0,3)$ & $B_iB_jB_k$ \\
 \hline
 \end{tabular}
 \caption{The building blocks of the tetrahedron (anti-tetrahedron) operators. Each of these operators satisfy the tetrahedron equation.}
 \label{tab:3simplexOP}
\end{table}
Linear combinations of the operators in a given type satisfy the tetrahedron equation. Linear combinations of the $(3,0)$ and the $(0,3)$ types satisfy the anti-tetrahedron equation (See Appendix \ref{app:antidsimplex}). Linear combinations of the $(3,0)$ and the $(1,2)$ types :
\begin{equation}\label{eq:Rd3mixed}
    R_{ijk} = \mu_0~A_iA_jA_k + \mu_1~B_iB_jA_k + \mu_2~B_iA_jB_k + \mu_3~A_iB_jB_k,
\end{equation} and of the $(0,3)$ and the $(2,1)$ types : 
\begin{equation}\label{eq:Rd3mixed-BA}
    R_{ijk} = \mu_0~B_iB_jB_k + \mu_1~A_iA_jB_k + \mu_2~A_iB_jA_k + \mu_3~B_iA_jA_k,
\end{equation} satisfy the vertex form of the tetrahedron equation \eqref{eq:d3simplex}. We notice that in each of these tetrahedron operators the constituents commute with each other. This pattern will continue for the higher simplex cases as we shall now see. The proofs for all these cases are similar to the previous ones and so we omit them.

Furthermore we could also consider linear combinations of the types $(3,0)$ [$(0,3)$] and $(2,1 )$ [$(1,2)$] or $(2,1)$ and $(1,2)$. These will satisfy the anti-tetrahedron equation (See Appendix \ref{app:antidsimplex}).
Finally we can consider linear combinations of three or all four types of operators. All of these satisfy a ``tetrahedron-like'' equation. We will not consider such cases in this paper.

\subsection{$4$ - Simplex or the Bazhanov-Stroganov's equation}
\label{subsec:d4}
Constructing the solutions for the 4-simplex equation will guide us in generalising the Clifford solutions to arbitrary even $d$. As in the previous cases we start with the simplest examples and systematically generalise them to more complicated ones by exploiting their linear structure.

Consider the following ansatzes for $R_{ijkl}$
\begin{equation}\label{eq:Rd4product-1}
    A_iA_jA_kA_l,~B_iB_jB_kB_l.
\end{equation}
These are operators of types $(4,0)$ and $(0,4)$ respectively.
They satisfy the constant 4-simplex equation
\begin{eqnarray}
     R_{1234}R_{1567}R_{2589}R_{368,10}R_{479,10} 
     =  R_{479,10}R_{368,10}R_{2589}R_{1567}R_{1234}, \label{eq:d4simplex}
\end{eqnarray}
as both sides simplify to 
$$ \prod\limits_{j=1}^{10}~A_j^2,~\prod\limits_{j=1}^{10}~B_j^2,$$
respectively.  Another set of non-trivial ansatzes for $R_{ijkl}$ is given by  
\begin{equation}\label{eq:Rd4product-2}
A_iA_jB_kB_l,~A_iB_jA_kB_l,~B_iA_jA_kB_l,~A_iB_jB_kA_l,~B_iA_jB_kA_l,~B_iB_jA_kA_l .
\end{equation}
These are operators of type $(2,2)$.
They satisfy the 4-simplex equation \eqref{eq:d4simplex} by reducing both sides of the equation to terms like
$$ A_1^2A_2^2\left(B_3A_3\right)\left(B_4A_4\right)A_5^2\left(B_6A_6\right)\left(B_7A_7\right)B_8^2B_9^2B_{10}^2.$$
We expect any linear combination of these two sets, \eqref{eq:Rd4product-1} and \eqref{eq:Rd4product-2}, to be solutions of the 4-simplex equation in \eqref{eq:d4simplex}. However to organise the solutions and simplify the upcoming proofs we will make a split among these operators. We consider two types of 4-simplex operators to show the linear structure :
\begin{eqnarray}
    R_{ijkl} & = & A_iA_jA_kA_l + B_iB_jB_kB_l \label{eq:R1d4}\\
    R_{ijkl} & = & A_iA_jB_kB_l + A_iB_jA_kB_l + B_iA_jA_kB_l \nonumber \\ & + & A_iB_jB_kA_l + B_iA_jB_kA_l + B_iB_jA_kA_l. \label{eq:R2d4}
\end{eqnarray}
The ansatz in \eqref{eq:R1d4} is a linear combination of the $(4,0)$ and the $(0,4)$ types whereas the second one in \eqref{eq:R2d4} is made of operators of the type $(2,2)$. The proofs that they solve \eqref{eq:d4simplex}, though straightforward, are once again cumbersome due to growing number of indices and terms. To simplify the proofs we will use notation introduced in \eqref{eq:Rd3shorthand} adapted to the 4-simplex operators. Thus the shorthand notations become
\begin{eqnarray}
    R_{ijkl}(\kappa_1,\kappa_2) & = & \kappa_1~A_iA_jA_kA_l + \kappa_2~B_iB_jB_kB_l \equiv \left(\kappa_1,\kappa_2\right)_{ijkl} \label{eq:R1d4shorthand}\\
    R_{ijkl}(\kappa_1,\cdots \kappa_6) & = & \kappa_1~A_iA_jB_kB_l + \kappa_2~A_iB_jA_kB_l + \kappa_3~B_iA_jA_kB_l \nonumber \\ & + & \kappa_4~A_iB_jB_kA_l + \kappa_5~B_iA_jB_kA_l + \kappa_6~B_iB_jA_kA_l \nonumber \\ & \equiv & \left(\kappa_1,\kappa_2,\kappa_3,\kappa_4,\kappa_5,\kappa_6\right)_{ijkl}.  \label{eq:R2d4shorthand}
\end{eqnarray}
Here the $\kappa_i$'s take values in $\left(+, -\right)$. As before, these parameters are not to be confused with spectral parameters. Using this notation the proofs that the 4-simplex operators \eqref{eq:R1d4} and \eqref{eq:R2d4} satisfy the 4-simplex equation \eqref{eq:d4simplex} are 
\begin{eqnarray}
  R_{1234}R_{1567}R_{2589}R_{368,10}R_{479,10}   & = & R_{479,10}\left(+, -\right)_{1234}\left(+, -\right)_{1567}\left(+, -\right)_{2589}\left(+, -\right)_{368,10} \nonumber \\
    & = & R_{479,10}R_{368,10}\left(+, +\right)_{1234}\left(+, +\right)_{1567}\left(+, +\right)_{2589} \nonumber \\
    & = & R_{479,10}R_{368,10}R_{2589}\left(+, -\right)_{1234}\left(+, -\right)_{1567} \nonumber \\
    & = & R_{479,10}R_{368,10}R_{2589}R_{1567}R_{1234},
\end{eqnarray}
and 
\begin{eqnarray}
  R_{1234}R_{1567}R_{2589}R_{368,10}R_{479,10}   & = & R_{479,10}\left(+,+,+,-,-,-\right)_{1234}\left(+,+,+,-,-,- \right)_{1567} \nonumber \\ 
    & \times & \left(+,+,+,-,-,-\right)_{2589}\left(+,+,+,-,-,-\right)_{368,10} \nonumber \\
    & = & R_{479,10}R_{368,10}\left(+,-,-,-,-,+\right)_{1234} \nonumber \\
    & \times & \left(+,-,-,-,-,+\right)_{1567}\left(+,-,-,-,-,+\right)_{2589}\nonumber \\
    & = & R_{479,10}R_{368,10}R_{2589} \nonumber \\
    & \times & \left(+,+,-,+,-,-\right)_{1234}\left(+,+,-,+,-,-\right)_{1567}\nonumber \\
    & = & R_{479,10}R_{368,10}R_{2589}R_{1567}R_{1234}
\end{eqnarray}
respectively.

As in the previous cases these solutions can be generalised to include arbitrary coefficients and more number of operators from the anticommuting sets of \eqref{eq:ABClifford}. The most general form of the 4-simplex operator in \eqref{eq:R1d4} is
\begin{eqnarray}\label{eq:R1d4general}
    R_{ijkl}\left(\bm{\mu}_{ijkl}, \bm{\nu}_{ijkl} \right) & = & \sum\limits_{m_1,m_2,m_3,m_4=1}^r~\left(\mu_{m_1m_2m_3m_4} \right)_{ijkl}\left(A^{(m_1)}\right)_i\left(A^{(m_2)}\right)_j \left(A^{(m_3)}\right)_k\left(A^{(m_4)}\right)_l \nonumber \\
    & + & \sum\limits_{n_1,n_2,n_3,n_4=1}^s~\left(\nu_{n_1n_2n_3n_4} \right)_{ijkl}\left(B^{(n_1)}\right)_i\left(B^{(n_2)}\right)_j \left(B^{(n_3)}\right)_k\left(B^{(n_4)}\right)_l,
\end{eqnarray}
where the $\mu$'s and $\nu$'s are spectral parameters.
This satisfies the spectral parameter dependent 4-simplex equation 
\begin{eqnarray}
    & & R_{1234}\left(\bm{\mu}_{1234}, \bm{\nu}_{1234} \right)R_{1567}\left(\bm{\mu}_{1567}, \bm{\nu}_{1567} \right)R_{2589}\left(\bm{\mu}_{2589}, \bm{\nu}_{2589} \right) \nonumber \\ & \times & R_{368,10}\left(\bm{\mu}_{368,10}, \bm{\nu}_{368,10} \right)R_{479,10}\left(\bm{\mu}_{479,10}, \bm{\nu}_{479,10} \right) \nonumber \\ 
    &  = & R_{479,10}\left(\bm{\mu}_{479,10}, \bm{\nu}_{479,10} \right)R_{368,10}\left(\bm{\mu}_{368,10}, \bm{\nu}_{368,10} \right)R_{2589}\left(\bm{\mu}_{2589}, \bm{\nu}_{2589} \right) \nonumber \\ & \times & R_{1567}\left(\bm{\mu}_{1567}, \bm{\nu}_{1567} \right)R_{1234}\left(\bm{\mu}_{1234}, \bm{\nu}_{1234} \right). \label{eq:d4simplexSPgeneral}
\end{eqnarray}
The generalised 4-simplex operators corresponding to \eqref{eq:R2d4} are quite tedious to write down due to the large number of indices on the coefficients. Nevertheless the expressions are rather straightforward when we follow the logic used in constructing the solutions in the previous cases. We hope this is clear to the reader by now and so we avoid writing down such long expressions in the rest of this paper. Two remarks are in order before we proceed to the higher simplex cases.

{\it {\bf Remark 1 :}} As the solutions are symmetric in the number of $A$'s and $B$'s in each term, their interchange leaves them invariant up to the coefficients multiplying each term. Nevertheless there could be local invertible operators\footnote{See also {\it {\bf Remark 1}} of Sec. \ref{subsec:d3}.} that relate the two $R$-matrices. For example for the $R$-matrices in \eqref{eq:R1d4general} we have 
\begin{equation}
    \Tilde{R}_{ijkl}\left(\bm{\mu}_{ijkl}, \bm{\nu}_{ijkl} \right) = \kappa~Q_iQ_jQ_kQ_l~R_{ijkl}\left(\bm{\mu}_{ijkl}, \bm{\nu}_{ijkl} \right)~Q_l^{-1}Q_k^{-1}Q_j^{-1}Q_i^{-1},
\end{equation}
with $\kappa$ as a complex number. As both $R$ and $\Tilde{R}$ solve the 4-simplex equation they are not taken to be different solutions if the above holds.


{\it {\bf Remark 2 :}} The 4-site operators that solve the 4-simplex equation are made up of an even number of $A$'s and $B$'s. We will see that this is the general requirement for the ansatzes to solve the appropriate higher simplex equations. For the $d=4$ case this leaves out operators of the form 
$$A_iA_jA_kB_l,~A_iA_jB_kA_l,~A_iB_jA_kA_l,~B_iA_jA_kA_l,$$
and those obtained by interchanging $A$ and $B$ in the above set. These are the operators of types $(3,1)$ and $(1,3)$ respectively. Linear combinations of these operators satisfy the anti-4-simplex equation discussed in Appendix \ref{app:antidsimplex}.

{\it {\bf Summary :}} We summarise the 4-simplex operators in Table \ref{tab:4simplexOP}.
\begin{table}[h!]
\centering
\begin{tabular}{ |c|p{8cm}| } 
 \hline
 Type & Operators  \\
 \hline
 \hline
 $(4,0)$ & $A_iA_jA_kA_l$  \\
 \hline
 $(3,1)$ & $A_iA_jA_kB_l$, $A_iA_jB_kA_l$, $A_iB_jA_kA_l$, $B_iA_jA_kA_l$ \\ 
 \hline
 $(2,2)$ & $A_iA_jB_kB_l$, $A_iB_jA_kB_l$, $B_iA_jA_kB_l$, \newline $A_iB_jB_kA_l$, $B_iA_jB_kA_l$, $B_iB_jA_kA_l$ \\
 \hline
 $(1,3)$ & $B_iB_jB_kA_l$, $B_iB_jA_kB_l$, $B_iA_jB_kB_l$, $A_iB_jB_kB_l$ \\
 \hline
 $(0,4)$ & $B_iB_jB_kB_l$  \\
 \hline
 \end{tabular}
 \caption{The building blocks of the 4-simplex (anti-4-simplex) operators.}
 \label{tab:4simplexOP}
\end{table}
In this case linear combinations of the types $(4,0)$, $(2,2)$ and $(0,4)$ satisfy the 4-simplex equation, whereas the linear combinations of the types $(3,1)$ and $(1,3)$ solve the anti-4-simplex equation.  All other possibilities solve a ``4-simplex-like'' equation which we do not consider in this paper.

\subsection{$5$ - Simplex equation}
\label{subsec:d5}
The logic to construct the $d$-simplex operators for odd $d$ will become clear when we write down the Clifford solutions for the 5-simplex equation. 
As always we begin with the simplest solutions, increasing the complexity in a linear manner. Following the pattern observed in the $d=2,3$ and 4 cases we make ansatzes with 5-site operators that have an even number of either the $A$ or $B$ operators. These include sets of types $(4,1)$ and $(3,2)$:
\begin{equation}\label{eq:Rd5product-1}
A_{i_1}A_{i_2}A_{i_3}A_{i_4}B_{i_5},~A_{i_1}A_{i_2}A_{i_3}B_{i_4}A_{i_5},~A_{i_1}A_{i_2}B_{i_3}A_{i_4}A_{i_5},~A_{i_1}B_{i_2}A_{i_3}A_{i_4}A_{i_5},~B_{i_1}A_{i_2}A_{i_3}A_{i_4}A_{i_5}, 
\end{equation}
and 
\begin{eqnarray}\label{eq:Rd5product-2}
    & & B_{i_1}B_{i_2}A_{i_3}A_{i_4}A_{i_5},~ B_{i_1}A_{i_2}B_{i_3}A_{i_4}A_{i_5},~ B_{i_1}A_{i_2}A_{i_3}B_{i_4}A_{i_5}, \nonumber \\ 
    & & B_{i_1}A_{i_2}A_{i_3}A_{i_4}B_{i_5},~ A_{i_1}B_{i_2}B_{i_3}A_{i_4}A_{i_5},~ A_{i_1}B_{i_2}A_{i_3}B_{i_4}A_{i_5}, \nonumber \\
    & & A_{i_1}B_{i_2}A_{i_3}A_{i_4}B_{i_5},~ A_{i_1}A_{i_2}B_{i_3}B_{i_4}A_{i_5},~ A_{i_1}A_{i_2}B_{i_3}A_{i_4}B_{i_5}, \nonumber \\
    & & A_{i_1}A_{i_2}A_{i_3}B_{i_4}B_{i_5},
\end{eqnarray}
respectively. 
It is easy to check that each of these operators satisfy the constant 5-simplex equation
\begin{eqnarray}
    & & R_{12345}R_{16789}R_{26,10,11,12}R_{37,10,13,14}R_{48,11,13,15}R_{59,12,14,15} \nonumber \\
    & = & R_{59,12,14,15}R_{48,11,13,15}R_{37,10,13,14}R_{26,10,11,12}R_{16789}R_{12345}. \nonumber \\ \label{eq:d5simplex}
\end{eqnarray} 
Linear combination of the operators within a given type is expected to satisfy the 5-simplex equation. Note that operators from these two sets anticommute with each other and hence ansatzes made of mixed types, $(4,1)$ and $(3,2)$, are not expected to solve the 5-simplex equation. Instead they solve the anti-5-simplex equation (See Appendix \ref{app:antidsimplex}). Using the operators in the sets \eqref{eq:Rd5product-1} and \eqref{eq:Rd5product-2} the two ansatzes  are
\begin{eqnarray}
    R_{i_1i_2i_3i_4i_5} & = & A_{i_1}A_{i_2}A_{i_3}A_{i_4}B_{i_5} + A_{i_1}A_{i_2}A_{i_3}B_{i_4}A_{i_5} + A_{i_1}A_{i_2}B_{i_3}A_{i_4}A_{i_5} \nonumber \\
    & + & A_{i_1}B_{i_2}A_{i_3}A_{i_4}A_{i_5} + B_{i_1}A_{i_2}A_{i_3}A_{i_4}A_{i_5} \label{eq:R1d5} \\
    R_{i_1i_2i_3i_4i_5} & = & B_{i_1}B_{i_2}A_{i_3}A_{i_4}A_{i_5} + B_{i_1}A_{i_2}B_{i_3}A_{i_4}A_{i_5} + B_{i_1}A_{i_2}A_{i_3}B_{i_4}A_{i_5} \nonumber \\ 
    & + & B_{i_1}A_{i_2}A_{i_3}A_{i_4}B_{i_5} + A_{i_1}B_{i_2}B_{i_3}A_{i_4}A_{i_5} + A_{i_1}B_{i_2}A_{i_3}B_{i_4}A_{i_5} \nonumber \\
    & + & A_{i_1}B_{i_2}A_{i_3}A_{i_4}B_{i_5} + A_{i_1}A_{i_2}B_{i_3}B_{i_4}A_{i_5} + A_{i_1}A_{i_2}B_{i_3}A_{i_4}B_{i_5} \nonumber \\
    & + & A_{i_1}A_{i_2}A_{i_3}B_{i_4}B_{i_5}. \label{eq:R2d5}
\end{eqnarray}
Next we present the proof that the 5-simplex operator in \eqref{eq:R1d5} satisfies \eqref{eq:d5simplex}. We again use a shorthand notation to keep the proofs tidy. The shorthand notations in \eqref{eq:Rd3shorthand} and \eqref{eq:R1d4shorthand} are appropriately modified. As it is straightforward we omit the details. Using this notation the proof goes as
\begin{eqnarray}
    & & R_{12345}R_{16789}R_{26,10,11,12}R_{37,10,13,14}R_{48,11,13,15}R_{59,12,14,15} \nonumber \\
    & = & R_{59,12,14,15}\left(+,-,-,-,-\right)_{12345}\left(+,-,-,-,-\right)_{16789} \nonumber \\
    & \times & \left(+,-,-,-,- \right)_{26,10,11,12}\left(+,-,-,-,- \right)_{37,10,13,14}\nonumber \\ 
    & \times &\left(+,-,-,-,-\right)_{48,11,13,15} \nonumber \\
    & = & R_{59,12,14,15}R_{48,11,13,15}\left(+,+,-,-,-\right)_{12345} \nonumber \\
    & \times & \left(+,+,-,-,-\right)_{16789}\left(+,+,-,-,-\right)_{26,10,11,12}\nonumber \\
    & \times & \left(+,+,-,-,-\right)_{37,10,13,14} \nonumber \\
    & = & R_{59,12,14,15}R_{48,11,13,15}R_{37,10,13,14} \nonumber \\
    & \times & \left(+,+,+,-,-\right)_{12345}\left(+,+,+,-,-\right)_{16789} \nonumber \\
    & \times & \left(+,+,+,-,-\right)_{26,10,11,12}\nonumber \\
    & = & R_{59,12,14,15}R_{48,11,13,15}R_{37,10,13,14}R_{26,10,11,12} \nonumber \\
    & \times & \left(+,+,+,+,-\right)_{12345}\left(+,+,+,+,-\right)_{16789}\nonumber \\
    & = & R_{59,12,14,15}R_{48,11,13,15}R_{37,10,13,14}R_{26,10,11,12}R_{16789}R_{12345}. 
\end{eqnarray}
A similar proof can be written down for the second 5-simplex operator in \eqref{eq:R2d5}. We omit this as the entire philosophy of these solutions is rather clear now. 

The generalisations of both the 5-simplex operators, \eqref{eq:R1d5} and \eqref{eq:R2d5} by including arbitrary linear combinations and more operators follow in the same manner as in the earlier cases. However writing them down is rather cumbersome. Hence we will skip this part with the understanding that they exist. These generalised solutions will satisfy a 5-simplex equation with spectral parameters.

The solutions obtained by interchanging the $A$ and $B$ operators in \eqref{eq:R1d5} and \eqref{eq:R2d5} may or may not be equivalent to each other by local invertible operators\footnote{See {\it {\bf Remark 1}} in Secs. \ref{subsec:d3} and \ref{subsec:d4}.}. These are made from operators of the $(2,3)$ and $(1,4)$ types. The operators $A_{i_1}A_{i_2}A_{i_3}A_{i_4}A_{i_5}$ and $B_{i_1}B_{i_2}B_{i_3}B_{i_4}B_{i_5}$, of the $(5,0)$ and $(0,5)$ types, also solve the 5-simplex equation trivially. However their linear combinations solve the anti-5-simplex equation as shown in Appendix \ref{app:antidsimplex}.

{\it {\bf Summary :}} The constituents of the 5-simplex solutions are summarised in Table \ref{tab:5simplexOP}.
\begin{table}[h!]
\centering
\begin{tabular}{ |c|p{8cm}| } 
 \hline
 Type & Operators  \\
 \hline
 \hline
 $(5,0)$ & $A_iA_jA_kA_lA_m$  \\
 \hline
 $(4,1)$ & $A_iA_jA_kA_lB_m$, $A_iA_jA_kB_lA_m$, $A_iA_jB_kA_lA_m$, $A_iB_jA_kA_lA_m$, $B_iA_jA_kA_lA_m$ \\ 
 \hline
 $(3,2)$ & $A_iA_jA_kB_lB_m$, $A_iA_jB_kA_lB_m$, $A_iB_jA_kA_lB_m$, $B_iA_jA_kA_lB_m$, $A_iA_jB_kB_lA_m$ \newline $A_iB_jA_kB_lA_m$, $B_iA_jA_kB_lA_m$, $A_iB_jB_kA_lA_m$, $B_iA_jB_kA_lA_m$, $B_iB_jA_kA_lA_m$ \\
 \hline
 $(2,3)$ & $B_iB_jB_kA_lA_m$, $B_iB_jA_kB_lA_m$, $B_iA_jB_kB_lA_m$, $A_iB_jB_kB_lA_m$, $B_iB_jA_kA_lB_m$ \newline $B_iA_jB_kA_lB_m$, $A_iB_jB_kA_lB_m$, $B_iA_jA_kB_lB_m$, $A_iB_jA_kB_lB_m$, $A_iA_jB_kB_lB_m$ \\
 \hline
 $(1,4)$ & $B_iB_jB_kB_lA_m$, $B_iB_jB_kA_lB_m$, $B_iB_jA_kB_lB_m$, $B_iA_jB_kB_lB_m$, $A_iB_jB_kB_lB_m$ \\ 
 \hline
 $(0,5)$ & $B_iB_jB_kB_lB_m$  \\
 \hline
 \end{tabular}
 \caption{The building blocks of the 5-simplex (anti-5-simplex) operators. Each of these operators satisfy the 5-simplex equation \eqref{eq:d5simplex}.}
 \label{tab:5simplexOP}
\end{table}
We have seen that linear combinations of the operators of the $(4,1)$ [$(3,2)$] type satisfy the 5-simplex equation. Adding to these we have linear combinations of the types $(4,1)$ [$(1,4)$], $(2,3)$ [$(3,2)$] and $(0,5)$ [$(5,0)$] also solve the 5-simplex equation of \eqref{eq:d5simplex}. Linear combinations of the operators of the $(5,0)$ and the $(0,5)$ types, $(5,0)$ [$(0,5)$] and $(4,1)$ [$(1,4)$] types,  $(5,0)$ [$(0,5)$] and $(2,3)$ [$(3,2)$] types satisfy the anti-5-simplex equation (See Appendix \ref{app:antidsimplex}). Other linear combinations of the different types solve a ```5-simplex-like'' equation which we do not discuss in this work.

\subsection{General $d$ - simplex equation}
\label{subsec:gend}
The solutions for the $d$-simplex equation for an arbitrary $d$ vary for odd and even $d$. This pattern can be understood from the solutions constructed so far. We will now identify this pattern and write the general result in two theorems.

In each of the examples we have seen that the $d$-simplex operator is a sum of terms, each of which has $d$ indices. Let the number of $A$ ($B$) operators in each term be $a$ ($b$). We have $$ a+b =d.$$ Then the individual terms of the $d$-simplex operator take the form :
\begin{equation}\label{eq:ABgend}
    A_{i_1}\cdots A_{i_a}B_{j_1}\cdots B_{j_b}~;~i_1,\cdots i_a, j_1\cdots j_b\in\{1,\cdots d\}.
\end{equation}
All the indices appearing in this operator are distinct from each other. We denote such operators as the pair $(a,b)$ or of type $(a,b)$. There are a total of $\binom{d}{a}=\binom{d}{b}$ such operators for a given pair. The pair $(b,a)$ is obtained by interchanging the $A$ and $B$ operators. When $A$ and $B$ are not related by a similarity transform this pair results in a $d$-simplex operator that is inequivalent to the $d$-simplex operator obtained from the pair $(a,b)$. The operators in \eqref{eq:ABgend} satisfy the $d$-simplex equation when at least one of $a$ and $b$ is even. That is we require an even number of either or both the $A$ or $B$ operators in each of these terms. When one of $a$ or $b$ is even then $d$ is odd. When they are both even then $d$ is even. Thus the solutions get classified by these pairs. Within a given type specified by a pair, the $\binom{d}{a}$ operators mutually commute and any linear combination of them is also a solution of the $d$-simplex equation.  This is the linearity of the solutions for a given pair $(a,b)$. 

The linearity among different pairs depends on whether $d$ is odd or even. The even case is simpler and so we discuss that first. In this case both $a$ and $b$ in any pair are even. There are $\frac{d}{2}+1$ such pairs. The product operators of the form \eqref{eq:ABgend} appearing in different pairs mutually commute with each other. Any linear combination of these continues to solve the $d$-simplex equation. Thus the linearity, for the case when $d$ is even, extends to all the pairs as well. 

When $d$ is odd the linear combinations of only certain pairs result in $d$-simplex operators. The prescription for choosing such pairs is as follows. The simplest case consists of two pairs $(a_1,b_1)$ and $(a_2,b_2)$. This will solve the $d$-simplex equation if $a_1-a_2$ or $b_1-b_2$ is even. If there are more than two pairs involved, then we require $a_i-a_j$ to be even for all $i$ and $j$. Here $i$ and $j$ index the pairs to be involved in the linear combination.  For example when $d=7$, a linear combinations of the pairs $(6,1)$ and $(2,5)$, the pairs $(5,2)$ and $(3,4)$ or the pairs $(0,7)$ and $(4,3)$ form valid 7-simplex operators. An example involving more than two pairs is given by $(6,1)$, $(2,5)$, $(0,7)$ and $(4,3)$. The other possibility in the odd $d$ case involves linear combinations of pairs $(a_i,b_i)$ with $a_i-a_j$ odd. It is easy to see that this is possible only for two pairs. These operators mutually anticommute with each other. Their linear combinations solve the anti-$d$-simplex equation (See Appendix \ref{app:antidsimplex}).

The last case to consider is when both $a$ and $b$ are odd. This occurs when $d$ is even. In this case the operators in \eqref{eq:ABgend} satisfy the anti-$d$-simplex equation (See Appendix \ref{app:antidsimplex}). Linear combination of these pairs also solve the anti-$d$-simplex equation.

Each of the $d$-simplex operators corresponding to an allowed pair $(a,b)$ can be generalised with the inclusion of arbitrary linear combinations and more number of operators from the anticommuting sets of operators in \eqref{eq:ABClifford}. They satisfy the spectral parameter dependent $d$-simplex equations. 

{\it {\bf Summary :}} The above discussion can be summarised into two theorems : the first one specifies the conditions for when the space of operators of the form \eqref{eq:ABgend} form $d$-simplex operators and the second one does the same for the anti-$d$-simplex operators.

\begin{theorem}[$d$-Simplex Operators]
{\it Consider the set of type $(a,b)$ operators \eqref{eq:ABgend} where $a,b\in\{0,1,\cdots, d \}$ and $a+b=d$. These operators satisfy the $d$-simplex equation when at least one of $a$ and $b$ is even. Furthermore this set forms a linear space of $d$-simplex solutions when the pairs $(a_i, b_i)$, with $i\in\{1,\cdots, d+1\}$, are such that $a_i-a_j$ is even for all pairs $i,j\in\{1,\cdots, d+1\}$. }
\end{theorem}

\begin{theorem}[Anti-$d$-Simplex Operators]
{\it The operators in \eqref{eq:ABgend} satisfy the anti-$d$-simplex equation when both $a$ and $b$ are odd. This happens when $d$ is even. Any linear combinations of such pairs are also anti-$d$-simplex solutions and hence the set of such operators form a linear space. When $d$ is odd all the product operators in \eqref{eq:ABgend} solve the $d$-simplex equation but only certain linear combinations solve the anti-$d$-simplex equation. These result from the combination of precisely two pairs $(a_1,b_1)$ and $(a_2,b_2)$, with $a_i,b_i\in\{0,1,\cdots, d\}$, such that $a_1-a_2$ is odd.} 
\end{theorem}

All other linear combinations of different types not covered by the above cases result in operators that solve equations resembling the $d$-simplex equations. We reserve the study of this situation for another paper.

\section{Examples : Qubit solutions}
\label{sec:qubitsolutions}
The $d$-simplex operators presented so far are representation independent. In this section we will choose $V=\mathbb{C}^2$ for the local Hilbert space and write down the qubit representations of the $d$-simplex operators. In what follows we will use a pair of anticommuting operators, $A$ and $B$ to construct the solutions. In other words we take $r=s=1$ in \eqref{eq:ABClifford}. The solutions below can be generalised to the case where there are two sets of mutually anticommuting operators as in \eqref{eq:ABClifford}. 

\paragraph{{\bf Case 1} :} The operators $A$ and $B$ satisfy the conditions stated in Case 1 of Sec. \ref{subsec:clifford}. That is they anticommute and both of them square to $\mathbb{1}$.
A simple qubit representation for $A$ and $B$ are
$$ A = X =\begin{pmatrix}
    0 & 1 \\ 1 & 0
\end{pmatrix}~;~B = Z =\begin{pmatrix}
    1 & 0 \\ 0 & -1
\end{pmatrix},$$
the first and third Pauli matrices respectively. In this case the operators $A$ and $B$ can be interchanged using the Hadamard gate :
$$ H = \frac{X+Z}{\sqrt{2}}~;~H^2=\mathbb{1}.$$ Applying this operator on every site gives us another $d$-simplex operator that is equivalent to the original $d$-simplex operator. As discussed in Secs. \ref{subsec:d3}, \ref{subsec:d4}, these solutions fall in the same equivalence class defined by such local invertible operators.
\begin{enumerate}
    \item {\it A 2-simplex or {\bf Yang-Baxter operator} :} The qubit representation of the operator in \eqref{eq:Rd2nonadditive} becomes \footnote{This form  was first mentioned to us by Kun Zhang \cite{Zhang_2024}. }
    \small
    \begin{equation}\label{eq:Rd2matrix}
        R(\mu_1, \mu_2) = \mu_1~X\otimes X + \mu_2~Z\otimes Z = \left(
\begin{array}{cccc}
 \mu _2 & 0 & 0 & \mu _1 \\
 0 & -\mu _2 & \mu _1 & 0 \\
 0 & \mu _1 & -\mu _2 & 0 \\
 \mu _1 & 0 & 0 & \mu _2 \\
\end{array}
\right). 
    \end{equation}
    \normalsize
    These are known as $X$-type solutions. Their entangling properties are studied in \cite{Padmanabhan_2021}. This solution falls in the $(1,4)$ class of Hietarinta's classification of constant Yang-Baxter solutions in the qubit case \cite{Hietarinta_1992} (See also Appendix A of \cite{Padmanabhan_2021}).
    \item {\it A 3-simplex or {\bf tetrahedron operator} :} The qubit representation of the tetrahedron operator in \eqref{eq:Rd3nonadditive} is
    \small
    \begin{eqnarray}\label{eq:Rd3matrix}
      &&  R(\mu_1, \mu_2, \mu_3)  =  \mu_1~X\otimes X\otimes Z + \mu_2~X\otimes Z\otimes X + \mu_3~Z\otimes X\otimes X \\
        & = &  \left(
\begin{array}{cccccccc}
 0 & 0 & 0 & \mu _3 & 0 & \mu _2 & \mu _1 & 0 \\
 0 & 0 & \mu _3 & 0 & \mu _2 & 0 & 0 & -\mu _1 \\
 0 & \mu _3 & 0 & 0 & \mu _1 & 0 & 0 & -\mu _2 \\
 \mu _3 & 0 & 0 & 0 & 0 & -\mu _1 & -\mu _2 & 0 \\
 0 & \mu _2 & \mu _1 & 0 & 0 & 0 & 0 & -\mu _3 \\
 \mu _2 & 0 & 0 & -\mu _1 & 0 & 0 & -\mu _3 & 0 \\
 \mu _1 & 0 & 0 & -\mu _2 & 0 & -\mu _3 & 0 & 0 \\
 0 & -\mu _1 & -\mu _2 & 0 & -\mu _3 & 0 & 0 & 0 \\
\end{array}
\right). \nonumber
        \end{eqnarray}
        \normalsize

\item {\it {\bf 4-simplex operators} :} The qubit representations of the $(4,0)$ type 4-simplex operator in \eqref{eq:R1d4} and the $(2,2)$ type 4-simplex operator in \eqref{eq:R2d4} are 
    \small
    \begin{eqnarray}\label{eq:R1d4matrix}
       && R(\mu_1, \mu_2)  =  \mu_1~X\otimes X\otimes X\otimes X + \mu_2~Z\otimes Z\otimes Z\otimes Z \\
        & = & 
        \left(
\begin{array}{cccccccccccccccc}
 \mu _2 & 0 & 0 & 0 & 0 & 0 & 0 & 0 & 0 & 0 & 0 & 0 & 0 & 0 & 0 & \mu _1 \\
 0 & -\mu _2 & 0 & 0 & 0 & 0 & 0 & 0 & 0 & 0 & 0 & 0 & 0 & 0 & \mu _1 & 0 \\
 0 & 0 & -\mu _2 & 0 & 0 & 0 & 0 & 0 & 0 & 0 & 0 & 0 & 0 & \mu _1 & 0 & 0 \\
 0 & 0 & 0 & \mu _2 & 0 & 0 & 0 & 0 & 0 & 0 & 0 & 0 & \mu _1 & 0 & 0 & 0 \\
 0 & 0 & 0 & 0 & -\mu _2 & 0 & 0 & 0 & 0 & 0 & 0 & \mu _1 & 0 & 0 & 0 & 0 \\
 0 & 0 & 0 & 0 & 0 & \mu _2 & 0 & 0 & 0 & 0 & \mu _1 & 0 & 0 & 0 & 0 & 0 \\
 0 & 0 & 0 & 0 & 0 & 0 & \mu _2 & 0 & 0 & \mu _1 & 0 & 0 & 0 & 0 & 0 & 0 \\
 0 & 0 & 0 & 0 & 0 & 0 & 0 & -\mu _2 & \mu _1 & 0 & 0 & 0 & 0 & 0 & 0 & 0 \\
 0 & 0 & 0 & 0 & 0 & 0 & 0 & \mu _1 & -\mu _2 & 0 & 0 & 0 & 0 & 0 & 0 & 0 \\
 0 & 0 & 0 & 0 & 0 & 0 & \mu _1 & 0 & 0 & \mu _2 & 0 & 0 & 0 & 0 & 0 & 0 \\
 0 & 0 & 0 & 0 & 0 & \mu _1 & 0 & 0 & 0 & 0 & \mu _2 & 0 & 0 & 0 & 0 & 0 \\
 0 & 0 & 0 & 0 & \mu _1 & 0 & 0 & 0 & 0 & 0 & 0 & -\mu _2 & 0 & 0 & 0 & 0 \\
 0 & 0 & 0 & \mu _1 & 0 & 0 & 0 & 0 & 0 & 0 & 0 & 0 & \mu _2 & 0 & 0 & 0 \\
 0 & 0 & \mu _1 & 0 & 0 & 0 & 0 & 0 & 0 & 0 & 0 & 0 & 0 & -\mu _2 & 0 & 0 \\
 0 & \mu _1 & 0 & 0 & 0 & 0 & 0 & 0 & 0 & 0 & 0 & 0 & 0 & 0 & -\mu _2 & 0 \\
 \mu _1 & 0 & 0 & 0 & 0 & 0 & 0 & 0 & 0 & 0 & 0 & 0 & 0 & 0 & 0 & \mu _2 \\
\end{array}
\right). \nonumber
    \end{eqnarray}
   \begin{eqnarray}\label{eq:R2d4matrix}
       & & R(\mu_1,\cdots \mu_6)  =  \mu_1~X\otimes X\otimes Z\otimes Z + \mu_2~X\otimes Z\otimes X\otimes Z + \mu_3~Z\otimes X\otimes X\otimes Z  \nonumber \\
        & + & \mu_4~X\otimes Z\otimes Z\otimes X + \mu_5~Z\otimes X\otimes Z\otimes X + \mu_6~Z\otimes Z\otimes X\otimes X \\
        & = & \left(
\begin{array}{cccccccccccccccc}
 0 & 0 & 0 & \mu _6 & 0 & \mu _5 & \mu _3 & 0 & 0 & \mu _4 & \mu _2 & 0 & \mu _1 & 0 & 0 & 0 \\
 0 & 0 & \mu _6 & 0 & \mu _5 & 0 & 0 & -\mu _3 & \mu _4 & 0 & 0 & -\mu _2 & 0 & -\mu _1 & 0 & 0 \\
 0 & \mu _6 & 0 & 0 & \mu _3 & 0 & 0 & -\mu _5 & \mu _2 & 0 & 0 & -\mu _4 & 0 & 0 & -\mu _1 & 0 \\
 \mu _6 & 0 & 0 & 0 & 0 & -\mu _3 & -\mu _5 & 0 & 0 & -\mu _2 & -\mu _4 & 0 & 0 & 0 & 0 & \mu _1 \\
 0 & \mu _5 & \mu _3 & 0 & 0 & 0 & 0 & -\mu _6 & \mu _1 & 0 & 0 & 0 & 0 & -\mu _4 & -\mu _2 & 0 \\
 \mu _5 & 0 & 0 & -\mu _3 & 0 & 0 & -\mu _6 & 0 & 0 & -\mu _1 & 0 & 0 & -\mu _4 & 0 & 0 & \mu _2 \\
 \mu _3 & 0 & 0 & -\mu _5 & 0 & -\mu _6 & 0 & 0 & 0 & 0 & -\mu _1 & 0 & -\mu _2 & 0 & 0 & \mu _4 \\
 0 & -\mu _3 & -\mu _5 & 0 & -\mu _6 & 0 & 0 & 0 & 0 & 0 & 0 & \mu _1 & 0 & \mu _2 & \mu _4 & 0 \\
 0 & \mu _4 & \mu _2 & 0 & \mu _1 & 0 & 0 & 0 & 0 & 0 & 0 & -\mu _6 & 0 & -\mu _5 & -\mu _3 & 0 \\
 \mu _4 & 0 & 0 & -\mu _2 & 0 & -\mu _1 & 0 & 0 & 0 & 0 & -\mu _6 & 0 & -\mu _5 & 0 & 0 & \mu _3 \\
 \mu _2 & 0 & 0 & -\mu _4 & 0 & 0 & -\mu _1 & 0 & 0 & -\mu _6 & 0 & 0 & -\mu _3 & 0 & 0 & \mu _5 \\
 0 & -\mu _2 & -\mu _4 & 0 & 0 & 0 & 0 & \mu _1 & -\mu _6 & 0 & 0 & 0 & 0 & \mu _3 & \mu _5 & 0 \\
 \mu _1 & 0 & 0 & 0 & 0 & -\mu _4 & -\mu _2 & 0 & 0 & -\mu _5 & -\mu _3 & 0 & 0 & 0 & 0 & \mu _6 \\
 0 & -\mu _1 & 0 & 0 & -\mu _4 & 0 & 0 & \mu _2 & -\mu _5 & 0 & 0 & \mu _3 & 0 & 0 & \mu _6 & 0 \\
 0 & 0 & -\mu _1 & 0 & -\mu _2 & 0 & 0 & \mu _4 & -\mu _3 & 0 & 0 & \mu _5 & 0 & \mu _6 & 0 & 0 \\
 0 & 0 & 0 & \mu _1 & 0 & \mu _2 & \mu _4 & 0 & 0 & \mu _3 & \mu _5 & 0 & \mu _6 & 0 & 0 & 0 \\
\end{array}
\right). \nonumber
    \end{eqnarray}
    \normalsize
Similar solutions occur as generalised Yang-Baxter operators \cite{Rowell2007ExtraspecialTG,Padmanabhan_2020}. 

\end{enumerate}


\paragraph{{\bf Case 2} :} The $A$ and $B$ operators now satisfy the conditions of the second case in Sec. \ref{subsec:clifford}. That is they anticommute and $A^2=0$ and $B^2=\mathbb{1}$. A possible qubit representation is given by :
$$ A = X\left(\frac{\mathbb{1}+Z}{2}\right)=\begin{pmatrix}
    0 & 0 \\ 1 & 0
\end{pmatrix}~;~B = Z = \begin{pmatrix}
    1 & 0 \\ 0 & -1
\end{pmatrix}.$$ These operators have unequal eigenvalues and hence cannot be related by a similarity transform. Thus for these choices of $A$ and $B$ we can obtain inequivalent $d$-simplex operators by interchanging them. We will explicitly see this in each of the examples below.

\begin{enumerate}
    \item {\it A 2-simplex or {\bf Yang-Baxter operator} :} The qubit representation of the operator in \eqref{eq:Rd2nonadditive} becomes 
    \small
    \begin{equation}\label{eq:Rd2matrix-Case2}
        R(\mu_1, \mu_2) = \mu_1~A\otimes A + \mu_2~Z\otimes Z = \left(
\begin{array}{cccc}
 \mu _2 & 0 & 0 & 0 \\
 0 & -\mu _2 & 0 & 0 \\
 0 & 0 & -\mu _2 & 0 \\
 \mu _1 & 0 & 0 & \mu _2 \\
\end{array}
\right)
    \end{equation}
    \normalsize
Interchanging $A$ and $Z$ in the expression \eqref{eq:Rd2matrix-Case2} amounts to interchanging the parameters $\mu_1$ and $\mu_2$. The eigenvalues of the two operators are $$ \{-\mu_1, -\mu_1, \mu_1, \mu_1 \}~;~\{-\mu_2, -\mu_2, \mu_2, \mu_2 \}$$ respectively. These operators fall into the $(0,1)$ class of Hietarinta's classification \cite{Hietarinta_1992} (See also Appendix A of \cite{Padmanabhan_2021}).  
    \item {\it A 3-simplex or {\bf tetrahedron operator} :} The qubit representation of the tetrahedron operators in \eqref{eq:Rd3nonadditive} and with $A$ and $Z$ interchanged are 
    \small
    \begin{eqnarray}\label{eq:Rd3matrix-Case2}
      &&  R(\mu_1, \mu_2, \mu_3)  =  \mu_1~A\otimes A\otimes Z + \mu_2~A\otimes Z\otimes A + \mu_3~Z\otimes A\otimes A  \\
        & = &  \left(
\begin{array}{cccccccc}
 0 & 0 & 0 & 0 & 0 & 0 & 0 & 0 \\
 0 & 0 & 0 & 0 & 0 & 0 & 0 & 0 \\
 0 & 0 & 0 & 0 & 0 & 0 & 0 & 0 \\
 \mu _3 & 0 & 0 & 0 & 0 & 0 & 0 & 0 \\
 0 & 0 & 0 & 0 & 0 & 0 & 0 & 0 \\
 \mu _2 & 0 & 0 & 0 & 0 & 0 & 0 & 0 \\
 \mu _1 & 0 & 0 & 0 & 0 & 0 & 0 & 0 \\
 0 & -\mu _1 & -\mu _2 & 0 & -\mu _3 & 0 & 0 & 0 \\
\end{array}
\right). \nonumber
\end{eqnarray}
\begin{eqnarray}\label{eq:Rd3matrix-Case2-BA}
      &&  R(\mu_1, \mu_2, \mu_3)  =  \mu_1~Z\otimes Z\otimes A + \mu_2~Z\otimes A\otimes Z + \mu_3~A\otimes Z\otimes Z \\
        & = &  \left(
\begin{array}{cccccccc}
 0 & 0 & 0 & 0 & 0 & 0 & 0 & 0 \\
 \mu _1 & 0 & 0 & 0 & 0 & 0 & 0 & 0 \\
 \mu _2 & 0 & 0 & 0 & 0 & 0 & 0 & 0 \\
 0 & -\mu _2 & -\mu _1 & 0 & 0 & 0 & 0 & 0 \\
 \mu _3 & 0 & 0 & 0 & 0 & 0 & 0 & 0 \\
 0 & -\mu _3 & 0 & 0 & -\mu _1 & 0 & 0 & 0 \\
 0 & 0 & -\mu _3 & 0 & -\mu _2 & 0 & 0 & 0 \\
 0 & 0 & 0 & \mu _3 & 0 & \mu _2 & \mu _1 & 0 \\
\end{array}
\right). \nonumber
\end{eqnarray}
\normalsize
These are non-invertible. The operator in \eqref{eq:Rd3matrix-Case2}  can be turned into an invertible operator by adding $Z\otimes Z\otimes Z$  :
 \begin{eqnarray}\label{eq:Rd3matrix-Case2-invertible}
      R(\mu_0, \mu_1, \mu_2, \mu_3)  & = & \mu_0~Z\otimes Z\otimes Z+ \mu_1~A\otimes A\otimes Z \nonumber \\ & + &  \mu_2~A\otimes Z\otimes A + \mu_3~Z\otimes A\otimes A   \\
        & = & \left(
\begin{array}{cccccccc}
 \mu _0 & 0 & 0 & 0 & 0 & 0 & 0 & 0 \\
 0 & -\mu _0 & 0 & 0 & 0 & 0 & 0 & 0 \\
 0 & 0 & -\mu _0 & 0 & 0 & 0 & 0 & 0 \\
 \mu _3 & 0 & 0 & \mu _0 & 0 & 0 & 0 & 0 \\
 0 & 0 & 0 & 0 & -\mu _0 & 0 & 0 & 0 \\
 \mu _2 & 0 & 0 & 0 & 0 & \mu _0 & 0 & 0 \\
 \mu _1 & 0 & 0 & 0 & 0 & 0 & \mu _0 & 0 \\
 0 & -\mu _1 & -\mu _2 & 0 & -\mu _3 & 0 & 0 & -\mu _0 \\
\end{array}
\right).
\end{eqnarray}
The resulting operator continues to satisfy the tetrahedron equation \eqref{eq:tetrahedronVertexForm} and is a linear combination of the types $(2,1)$ and $(0,3)$.

\item {\it {\bf 4-simplex operators} :} The qubit representation of the $(4,0)$ type 4-simplex operator in \eqref{eq:R1d4} is 
    \small
    \begin{eqnarray}\label{eq:R1d4matrix-Case2}
       && R(\mu_1, \mu_2)  =  \mu_1~A\otimes A\otimes A\otimes A + \mu_2~Z\otimes Z\otimes Z\otimes Z \\
        & = & \left(
\begin{array}{cccccccccccccccc}
 \mu _2 & 0 & 0 & 0 & 0 & 0 & 0 & 0 & 0 & 0 & 0 & 0 & 0 & 0 & 0 & 0 \\
 0 & -\mu _2 & 0 & 0 & 0 & 0 & 0 & 0 & 0 & 0 & 0 & 0 & 0 & 0 & 0 & 0 \\
 0 & 0 & -\mu _2 & 0 & 0 & 0 & 0 & 0 & 0 & 0 & 0 & 0 & 0 & 0 & 0 & 0 \\
 0 & 0 & 0 & \mu _2 & 0 & 0 & 0 & 0 & 0 & 0 & 0 & 0 & 0 & 0 & 0 & 0 \\
 0 & 0 & 0 & 0 & -\mu _2 & 0 & 0 & 0 & 0 & 0 & 0 & 0 & 0 & 0 & 0 & 0 \\
 0 & 0 & 0 & 0 & 0 & \mu _2 & 0 & 0 & 0 & 0 & 0 & 0 & 0 & 0 & 0 & 0 \\
 0 & 0 & 0 & 0 & 0 & 0 & \mu _2 & 0 & 0 & 0 & 0 & 0 & 0 & 0 & 0 & 0 \\
 0 & 0 & 0 & 0 & 0 & 0 & 0 & -\mu _2 & 0 & 0 & 0 & 0 & 0 & 0 & 0 & 0 \\
 0 & 0 & 0 & 0 & 0 & 0 & 0 & 0 & -\mu _2 & 0 & 0 & 0 & 0 & 0 & 0 & 0 \\
 0 & 0 & 0 & 0 & 0 & 0 & 0 & 0 & 0 & \mu _2 & 0 & 0 & 0 & 0 & 0 & 0 \\
 0 & 0 & 0 & 0 & 0 & 0 & 0 & 0 & 0 & 0 & \mu _2 & 0 & 0 & 0 & 0 & 0 \\
 0 & 0 & 0 & 0 & 0 & 0 & 0 & 0 & 0 & 0 & 0 & -\mu _2 & 0 & 0 & 0 & 0 \\
 0 & 0 & 0 & 0 & 0 & 0 & 0 & 0 & 0 & 0 & 0 & 0 & \mu _2 & 0 & 0 & 0 \\
 0 & 0 & 0 & 0 & 0 & 0 & 0 & 0 & 0 & 0 & 0 & 0 & 0 & -\mu _2 & 0 & 0 \\
 0 & 0 & 0 & 0 & 0 & 0 & 0 & 0 & 0 & 0 & 0 & 0 & 0 & 0 & -\mu _2 & 0 \\
 \mu _1 & 0 & 0 & 0 & 0 & 0 & 0 & 0 & 0 & 0 & 0 & 0 & 0 & 0 & 0 & \mu _2 \\
\end{array}
\right)
         \nonumber
    \end{eqnarray}
    \normalsize
Interchanging $A$ and $Z$ in \eqref{eq:R1d4matrix-Case2} results in an inequivalent operator with $\mu_1$ and $\mu_2$ interchanged. The qubit representation of the $(2,2)$ type 4-simplex operator in \eqref{eq:R2d4} is
    \small
    \begin{eqnarray}\label{eq:R2d4matrix-Case2noninvertible}
       & & R(\mu_1,\cdots \mu_6)  =  \mu_1~A\otimes A\otimes Z\otimes Z + \mu_2~A\otimes Z\otimes A\otimes Z + \mu_3~Z\otimes A\otimes A\otimes Z  \nonumber \\
        & + & \mu_4~A\otimes Z\otimes Z\otimes A + \mu_5~Z\otimes A\otimes Z\otimes A + \mu_6~Z\otimes Z\otimes A\otimes A \\
        & = & \left(
\begin{array}{cccccccccccccccc}
 0 & 0 & 0 & 0 & 0 & 0 & 0 & 0 & 0 & 0 & 0 & 0 & 0 & 0 & 0 & 0 \\
 0 & 0 & 0 & 0 & 0 & 0 & 0 & 0 & 0 & 0 & 0 & 0 & 0 & 0 & 0 & 0 \\
 0 & 0 & 0 & 0 & 0 & 0 & 0 & 0 & 0 & 0 & 0 & 0 & 0 & 0 & 0 & 0 \\
 \mu _6 & 0 & 0 & 0 & 0 & 0 & 0 & 0 & 0 & 0 & 0 & 0 & 0 & 0 & 0 & 0 \\
 0 & 0 & 0 & 0 & 0 & 0 & 0 & 0 & 0 & 0 & 0 & 0 & 0 & 0 & 0 & 0 \\
 \mu _5 & 0 & 0 & 0 & 0 & 0 & 0 & 0 & 0 & 0 & 0 & 0 & 0 & 0 & 0 & 0 \\
 \mu _3 & 0 & 0 & 0 & 0 & 0 & 0 & 0 & 0 & 0 & 0 & 0 & 0 & 0 & 0 & 0 \\
 0 & -\mu _3 & -\mu _5 & 0 & -\mu _6 & 0 & 0 & 0 & 0 & 0 & 0 & 0 & 0 & 0 & 0 & 0 \\
 0 & 0 & 0 & 0 & 0 & 0 & 0 & 0 & 0 & 0 & 0 & 0 & 0 & 0 & 0 & 0 \\
 \mu _4 & 0 & 0 & 0 & 0 & 0 & 0 & 0 & 0 & 0 & 0 & 0 & 0 & 0 & 0 & 0 \\
 \mu _2 & 0 & 0 & 0 & 0 & 0 & 0 & 0 & 0 & 0 & 0 & 0 & 0 & 0 & 0 & 0 \\
 0 & -\mu _2 & -\mu _4 & 0 & 0 & 0 & 0 & 0 & -\mu _6 & 0 & 0 & 0 & 0 & 0 & 0 & 0 \\
 \mu _1 & 0 & 0 & 0 & 0 & 0 & 0 & 0 & 0 & 0 & 0 & 0 & 0 & 0 & 0 & 0 \\
 0 & -\mu _1 & 0 & 0 & -\mu _4 & 0 & 0 & 0 & -\mu _5 & 0 & 0 & 0 & 0 & 0 & 0 & 0 \\
 0 & 0 & -\mu _1 & 0 & -\mu _2 & 0 & 0 & 0 & -\mu _3 & 0 & 0 & 0 & 0 & 0 & 0 & 0 \\
 0 & 0 & 0 & \mu _1 & 0 & \mu _2 & \mu _4 & 0 & 0 & \mu _3 & \mu _5 & 0 & \mu _6 & 0 & 0 & 0 \\
\end{array}
\right). \nonumber
    \end{eqnarray}
    \normalsize
Interchanging $A$ and $Z$ in \eqref{eq:R2d4matrix-Case2noninvertible} results in an inequivalent operator with the $\mu$'s shuffled. 

The $(2,2)$ type 4-simplex operator is not invertible while the $(4,0)$ type is invertible. We find that a linear combination of the two operators is invertible\footnote{Note that the linear combination is still a solution of the 4-simplex equation.} :  
 \begin{eqnarray}\label{eq:R2d4matrix-Case2invertible}
       & & R(\mu_1,\cdots \mu_6~;~\nu_1, \nu_2) \nonumber \\  &  = &   \mu_1~A\otimes A\otimes Z\otimes Z + \mu_2~A\otimes Z\otimes A\otimes Z + \mu_3~Z\otimes A\otimes A\otimes Z  \nonumber \\
        & + & \mu_4~A\otimes Z\otimes Z\otimes A + \mu_5~Z\otimes A\otimes Z\otimes A + \mu_6~Z\otimes Z\otimes A\otimes A  \nonumber \\
        & + & \nu_1~A\otimes A\otimes A\otimes A + \nu_2~Z\otimes Z\otimes Z\otimes Z.  
        \end{eqnarray}   

\end{enumerate}

   \paragraph{{\bf Case 3} :} Each of the operators $A$ and $B$ is either nilpotent or a projector, satisfying the conditions of Case 3 in Sec. \ref{subsec:clifford}. It turns out that the resulting $d$-simplex operators for $d=2,3,4$ cases are non-invertible in the qubit representation. However, in each of these cases, $AB=BA=0$. Thus we can add an identity operator into the different ansatzes and make them invertible. We will see this for the cases when both $A$ and $B$ are nilpotent and when they are both projectors (See Examples 2 and 3 of Case 3 in Sec. \ref{subsec:clifford}). 

   Consider the case when both are nilpotent :
   \begin{equation}\label{eq:ABCase3qubitreps-nilpotent}
       A = X\left(\frac{\mathbb{1}+Z}{2}\right)~;~B = Y\left(\frac{\mathbb{1}+Z}{2}\right).
   \end{equation}

   \begin{enumerate}
    \item {\it A 2-simplex or {\bf Yang-Baxter operator} :} Using these nilpotent operators, the $R$-matrix in \eqref{eq:Rd2nonadditive} becomes 
    \small
    \begin{equation}\label{eq:Rd2matrix-Case3}
        R(\mu_1, \mu_2) = \mu_0~\mathbb{1}\otimes\mathbb{1} + \mu_1~A\otimes A + \mu_2~B\otimes B = \left(
\begin{array}{cccc}
 \mu _0 & 0 & 0 & 0 \\
 0 & \mu _0 & 0 & 0 \\
 0 & 0 & \mu _0 & 0 \\
 \mu _1-\mu _2 & 0 & 0 & \mu _0 \\
\end{array}
\right).
\end{equation}
This solution falls in the $(2,3)$ class of Hietarinta's classification \cite{Hietarinta_1992} (See also Appendix A of \cite{Padmanabhan_2021}).
 \item {\it A 3-simplex or {\bf tetrahedron operator} :} The qubit representation of the tetrahedron operators in \eqref{eq:Rd3nonadditive} with $A$ and $B$ nilpotent are :
 \begin{eqnarray}\label{eq:Rd3matrix-Case3}
    & & R(\mu_0,\cdots, \mu_3) = \mu_0~\mathbb{1}\otimes\mathbb{1}\otimes\mathbb{1} + \mu_1~A\otimes B\otimes B + \mu_2~B\otimes A\otimes B + \mu_3~B\otimes B\otimes A  \nonumber \\
    & = & \left(
\begin{array}{cccccccc}
 \mu _0 & 0 & 0 & 0 & 0 & 0 & 0 & 0 \\
 0 & \mu _0 & 0 & 0 & 0 & 0 & 0 & 0 \\
 0 & 0 & \mu _0 & 0 & 0 & 0 & 0 & 0 \\
 0 & 0 & 0 & \mu _0 & 0 & 0 & 0 & 0 \\
 0 & 0 & 0 & 0 & \mu _0 & 0 & 0 & 0 \\
 0 & 0 & 0 & 0 & 0 & \mu _0 & 0 & 0 \\
 0 & 0 & 0 & 0 & 0 & 0 & \mu _0 & 0 \\
 -\mu _1-\mu _2-\mu _3 & 0 & 0 & 0 & 0 & 0 & 0 & \mu _0 \\
\end{array}
\right),
 \end{eqnarray}
 and 
 \begin{eqnarray}\label{eq:Rd3matrix-Case3-BA}
    & & R(\mu_0,\cdots, \mu_3) = \mu_0~\mathbb{1}\otimes\mathbb{1}\otimes\mathbb{1} + \mu_1~B\otimes A\otimes A + \mu_2~A\otimes B\otimes A + \mu_3~A\otimes A\otimes B  \nonumber \\
    & = & \left(
\begin{array}{cccccccc}
 \mu _0 & 0 & 0 & 0 & 0 & 0 & 0 & 0 \\
 0 & \mu _0 & 0 & 0 & 0 & 0 & 0 & 0 \\
 0 & 0 & \mu _0 & 0 & 0 & 0 & 0 & 0 \\
 0 & 0 & 0 & \mu _0 & 0 & 0 & 0 & 0 \\
 0 & 0 & 0 & 0 & \mu _0 & 0 & 0 & 0 \\
 0 & 0 & 0 & 0 & 0 & \mu _0 & 0 & 0 \\
 0 & 0 & 0 & 0 & 0 & 0 & \mu _0 & 0 \\
 i \left(\mu _1+\mu _2+\mu _3\right) & 0 & 0 & 0 & 0 & 0 & 0 & \mu _0 \\
\end{array}
\right).
    \end{eqnarray}
\item {\it {\bf 4-simplex operators} :} The invertible $(4,0)$ type 4-simplex operator in \eqref{eq:R1d4} is 
    \small
    \begin{eqnarray}\label{eq:R1d4matrix-Case3}
       && R(\mu_1, \mu_2)  =  \mu_0~\mathbb{1}\otimes\mathbb{1}\otimes\mathbb{1}\otimes\mathbb{1} + \mu_1~A\otimes A\otimes A\otimes A + \mu_2~B\otimes B\otimes B\otimes B \\
       & = & \left(
\begin{array}{cccccccccccccccc}
 \mu _0 & 0 & 0 & 0 & 0 & 0 & 0 & 0 & 0 & 0 & 0 & 0 & 0 & 0 & 0 & 0 \\
 0 & \mu _0 & 0 & 0 & 0 & 0 & 0 & 0 & 0 & 0 & 0 & 0 & 0 & 0 & 0 & 0 \\
 0 & 0 & \mu _0 & 0 & 0 & 0 & 0 & 0 & 0 & 0 & 0 & 0 & 0 & 0 & 0 & 0 \\
 0 & 0 & 0 & \mu _0 & 0 & 0 & 0 & 0 & 0 & 0 & 0 & 0 & 0 & 0 & 0 & 0 \\
 0 & 0 & 0 & 0 & \mu _0 & 0 & 0 & 0 & 0 & 0 & 0 & 0 & 0 & 0 & 0 & 0 \\
 0 & 0 & 0 & 0 & 0 & \mu _0 & 0 & 0 & 0 & 0 & 0 & 0 & 0 & 0 & 0 & 0 \\
 0 & 0 & 0 & 0 & 0 & 0 & \mu _0 & 0 & 0 & 0 & 0 & 0 & 0 & 0 & 0 & 0 \\
 0 & 0 & 0 & 0 & 0 & 0 & 0 & \mu _0 & 0 & 0 & 0 & 0 & 0 & 0 & 0 & 0 \\
 0 & 0 & 0 & 0 & 0 & 0 & 0 & 0 & \mu _0 & 0 & 0 & 0 & 0 & 0 & 0 & 0 \\
 0 & 0 & 0 & 0 & 0 & 0 & 0 & 0 & 0 & \mu _0 & 0 & 0 & 0 & 0 & 0 & 0 \\
 0 & 0 & 0 & 0 & 0 & 0 & 0 & 0 & 0 & 0 & \mu _0 & 0 & 0 & 0 & 0 & 0 \\
 0 & 0 & 0 & 0 & 0 & 0 & 0 & 0 & 0 & 0 & 0 & \mu _0 & 0 & 0 & 0 & 0 \\
 0 & 0 & 0 & 0 & 0 & 0 & 0 & 0 & 0 & 0 & 0 & 0 & \mu _0 & 0 & 0 & 0 \\
 0 & 0 & 0 & 0 & 0 & 0 & 0 & 0 & 0 & 0 & 0 & 0 & 0 & \mu _0 & 0 & 0 \\
 0 & 0 & 0 & 0 & 0 & 0 & 0 & 0 & 0 & 0 & 0 & 0 & 0 & 0 & \mu _0 & 0 \\
 \mu _1+\mu _2 & 0 & 0 & 0 & 0 & 0 & 0 & 0 & 0 & 0 & 0 & 0 & 0 & 0 & 0 & \mu _0 \\
\end{array}
\right).
       \end{eqnarray}
       In this representation the invertible 4-simplex operator of type $(2,2)$ is given by :
        \begin{eqnarray}\label{eq:R2d4matrix-Case3}
       && R(\mu_0,\cdots \mu_6)  =  \mu_0~\mathbb{1}\otimes\mathbb{1}\otimes\mathbb{1}\otimes\mathbb{1} + \mu_1~A\otimes A\otimes B\otimes B \nonumber \\ 
       & + & \mu_2~A\otimes B\otimes A\otimes B + \mu_3~B\otimes A\otimes A\otimes B + \mu_4~A\otimes B\otimes B\otimes A\nonumber \\ 
       & + & \mu_5~B\otimes A\otimes B\otimes A + \mu_6~B\otimes B\otimes A\otimes A \\
       & = & \left(
\begin{array}{cccccccccccccccc}
 \mu _0 & 0 & 0 & 0 & 0 & 0 & 0 & 0 & 0 & 0 & 0 & 0 & 0 & 0 & 0 & 0 \\
 0 & \mu _0 & 0 & 0 & 0 & 0 & 0 & 0 & 0 & 0 & 0 & 0 & 0 & 0 & 0 & 0 \\
 0 & 0 & \mu _0 & 0 & 0 & 0 & 0 & 0 & 0 & 0 & 0 & 0 & 0 & 0 & 0 & 0 \\
 0 & 0 & 0 & \mu _0 & 0 & 0 & 0 & 0 & 0 & 0 & 0 & 0 & 0 & 0 & 0 & 0 \\
 0 & 0 & 0 & 0 & \mu _0 & 0 & 0 & 0 & 0 & 0 & 0 & 0 & 0 & 0 & 0 & 0 \\
 0 & 0 & 0 & 0 & 0 & \mu _0 & 0 & 0 & 0 & 0 & 0 & 0 & 0 & 0 & 0 & 0 \\
 0 & 0 & 0 & 0 & 0 & 0 & \mu _0 & 0 & 0 & 0 & 0 & 0 & 0 & 0 & 0 & 0 \\
 0 & 0 & 0 & 0 & 0 & 0 & 0 & \mu _0 & 0 & 0 & 0 & 0 & 0 & 0 & 0 & 0 \\
 0 & 0 & 0 & 0 & 0 & 0 & 0 & 0 & \mu _0 & 0 & 0 & 0 & 0 & 0 & 0 & 0 \\
 0 & 0 & 0 & 0 & 0 & 0 & 0 & 0 & 0 & \mu _0 & 0 & 0 & 0 & 0 & 0 & 0 \\
 0 & 0 & 0 & 0 & 0 & 0 & 0 & 0 & 0 & 0 & \mu _0 & 0 & 0 & 0 & 0 & 0 \\
 0 & 0 & 0 & 0 & 0 & 0 & 0 & 0 & 0 & 0 & 0 & \mu _0 & 0 & 0 & 0 & 0 \\
 0 & 0 & 0 & 0 & 0 & 0 & 0 & 0 & 0 & 0 & 0 & 0 & \mu _0 & 0 & 0 & 0 \\
 0 & 0 & 0 & 0 & 0 & 0 & 0 & 0 & 0 & 0 & 0 & 0 & 0 & \mu _0 & 0 & 0 \\
 0 & 0 & 0 & 0 & 0 & 0 & 0 & 0 & 0 & 0 & 0 & 0 & 0 & 0 & \mu _0 & 0 \\
 -\sum\limits_{j=1}^6~\mu _j & 0 & 0 & 0 & 0 & 0 & 0 & 0 & 0 & 0 & 0 & 0 & 0 & 0 & 0 & \mu _0 \\
\end{array}
\right). \nonumber
       \end{eqnarray}
    \end{enumerate}

Next we consider the case when both $A$ and $B$ are projectors :
\begin{equation}\label{eq:ABCase3qubitreps-projectors}
    A = \left(\frac{\mathbb{1}+Z}{2}\right)~;~B = \left(\frac{\mathbb{1}-Z}{2}\right).
\end{equation}

\begin{enumerate}
    \item {\it A 2-simplex or {\bf Yang-Baxter operator} :} Using these projectors, the $R$-matrix in \eqref{eq:Rd2nonadditive} becomes 
    \small
    \begin{equation}\label{eq:Rd2matrix-Case3-projectors}
        R(\mu_1, \mu_2) = \mu_0~\mathbb{1}\otimes\mathbb{1} + \mu_1~A\otimes A + \mu_2~B\otimes B = \left(
\begin{array}{cccc}
 \mu _0+\mu _1 & 0 & 0 & 0 \\
 0 & \mu _0 & 0 & 0 \\
 0 & 0 & \mu _0 & 0 \\
 0 & 0 & 0 & \mu _0+\mu _2 \\
\end{array}
\right).
\end{equation}
This falls into the $(3,1)$ class of Hietarinta's classification of qubit $R$-matrices \cite{Hietarinta_1992} (See also Appendix A of \cite{Padmanabhan_2021}).
\item {\it A 3-simplex or {\bf tetrahedron operator} :} The qubit representation of the tetrahedron operators in \eqref{eq:Rd3nonadditive} with $A$ and $B$ as orthogonal projectors are :
 \begin{eqnarray}\label{eq:Rd3matrix-Case3-projectors}
    & & R(\mu_0,\cdots, \mu_3) = \mu_0~\mathbb{1}\otimes\mathbb{1}\otimes\mathbb{1} + \mu_1~A\otimes B\otimes B + \mu_2~B\otimes A\otimes B + \mu_3~B\otimes B\otimes A  \nonumber \\
    & = & \left(
\begin{array}{cccccccc}
 \mu _0 & 0 & 0 & 0 & 0 & 0 & 0 & 0 \\
 0 & \mu _0 & 0 & 0 & 0 & 0 & 0 & 0 \\
 0 & 0 & \mu _0 & 0 & 0 & 0 & 0 & 0 \\
 0 & 0 & 0 & \mu _0+\mu _1 & 0 & 0 & 0 & 0 \\
 0 & 0 & 0 & 0 & \mu _0 & 0 & 0 & 0 \\
 0 & 0 & 0 & 0 & 0 & \mu _0+\mu _2 & 0 & 0 \\
 0 & 0 & 0 & 0 & 0 & 0 & \mu _0+\mu _3 & 0 \\
 0 & 0 & 0 & 0 & 0 & 0 & 0 & \mu _0 \\
\end{array}
\right)
\end{eqnarray}
and 
\begin{eqnarray}\label{eq:Rd3matrix-Case3-BA-projectors}
    & & R(\mu_0,\cdots, \mu_3) = \mu_0~\mathbb{1}\otimes\mathbb{1}\otimes\mathbb{1} + \mu_1~B\otimes A\otimes A + \mu_2~A\otimes B\otimes A + \mu_3~A\otimes A\otimes B  \nonumber \\
    & = & \left(
\begin{array}{cccccccc}
 \mu _0 & 0 & 0 & 0 & 0 & 0 & 0 & 0 \\
 0 & \mu _0+\mu _3 & 0 & 0 & 0 & 0 & 0 & 0 \\
 0 & 0 & \mu _0+\mu _2 & 0 & 0 & 0 & 0 & 0 \\
 0 & 0 & 0 & \mu _0 & 0 & 0 & 0 & 0 \\
 0 & 0 & 0 & 0 & \mu _0+\mu _1 & 0 & 0 & 0 \\
 0 & 0 & 0 & 0 & 0 & \mu _0 & 0 & 0 \\
 0 & 0 & 0 & 0 & 0 & 0 & \mu _0 & 0 \\
 0 & 0 & 0 & 0 & 0 & 0 & 0 & \mu _0 \\
\end{array}
\right).
\end{eqnarray}
\item {\it {\bf 4-simplex operators} :} The invertible $(4,0)$ type 4-simplex operator in \eqref{eq:R1d4} is 
    \small
    \begin{eqnarray}\label{eq:R1d4matrix-Case3-projectors}
       && R(\mu_1, \mu_2)  =  \mu_0~\mathbb{1}\otimes\mathbb{1}\otimes\mathbb{1}\otimes\mathbb{1} + \mu_1~A\otimes A\otimes A\otimes A + \mu_2~B\otimes B\otimes B\otimes B \\
       & = & \left(
\begin{array}{cccccccccccccccc}
 \mu _0+\mu _1 & 0 & 0 & 0 & 0 & 0 & 0 & 0 & 0 & 0 & 0 & 0 & 0 & 0 & 0 & 0 \\
 0 & \mu _0 & 0 & 0 & 0 & 0 & 0 & 0 & 0 & 0 & 0 & 0 & 0 & 0 & 0 & 0 \\
 0 & 0 & \mu _0 & 0 & 0 & 0 & 0 & 0 & 0 & 0 & 0 & 0 & 0 & 0 & 0 & 0 \\
 0 & 0 & 0 & \mu _0 & 0 & 0 & 0 & 0 & 0 & 0 & 0 & 0 & 0 & 0 & 0 & 0 \\
 0 & 0 & 0 & 0 & \mu _0 & 0 & 0 & 0 & 0 & 0 & 0 & 0 & 0 & 0 & 0 & 0 \\
 0 & 0 & 0 & 0 & 0 & \mu _0 & 0 & 0 & 0 & 0 & 0 & 0 & 0 & 0 & 0 & 0 \\
 0 & 0 & 0 & 0 & 0 & 0 & \mu _0 & 0 & 0 & 0 & 0 & 0 & 0 & 0 & 0 & 0 \\
 0 & 0 & 0 & 0 & 0 & 0 & 0 & \mu _0 & 0 & 0 & 0 & 0 & 0 & 0 & 0 & 0 \\
 0 & 0 & 0 & 0 & 0 & 0 & 0 & 0 & \mu _0 & 0 & 0 & 0 & 0 & 0 & 0 & 0 \\
 0 & 0 & 0 & 0 & 0 & 0 & 0 & 0 & 0 & \mu _0 & 0 & 0 & 0 & 0 & 0 & 0 \\
 0 & 0 & 0 & 0 & 0 & 0 & 0 & 0 & 0 & 0 & \mu _0 & 0 & 0 & 0 & 0 & 0 \\
 0 & 0 & 0 & 0 & 0 & 0 & 0 & 0 & 0 & 0 & 0 & \mu _0 & 0 & 0 & 0 & 0 \\
 0 & 0 & 0 & 0 & 0 & 0 & 0 & 0 & 0 & 0 & 0 & 0 & \mu _0 & 0 & 0 & 0 \\
 0 & 0 & 0 & 0 & 0 & 0 & 0 & 0 & 0 & 0 & 0 & 0 & 0 & \mu _0 & 0 & 0 \\
 0 & 0 & 0 & 0 & 0 & 0 & 0 & 0 & 0 & 0 & 0 & 0 & 0 & 0 & \mu _0 & 0 \\
 0 & 0 & 0 & 0 & 0 & 0 & 0 & 0 & 0 & 0 & 0 & 0 & 0 & 0 & 0 & \mu _0+\mu _2 \\
\end{array}
\right).
       \end{eqnarray}
        In this representation the invertible 4-simplex operator of type $(2,2)$ is given by :
        \begin{eqnarray}\label{eq:R2d4matrix-Case3-projectors}
       && R(\mu_0,\cdots \mu_6)  =  \mu_0~\mathbb{1}\otimes\mathbb{1}\otimes\mathbb{1}\otimes\mathbb{1} + \mu_1~A\otimes A\otimes B\otimes B \nonumber \\ 
       & + & \mu_2~A\otimes B\otimes A\otimes B + \mu_3~B\otimes A\otimes A\otimes B + \mu_4~A\otimes B\otimes B\otimes A\nonumber \\ 
       & + & \mu_5~B\otimes A\otimes B\otimes A + \mu_6~B\otimes B\otimes A\otimes A \\
       & = & \left(
\begin{array}{cccccccccccccccc}
 \mu _0 & 0 & 0 & 0 & 0 & 0 & 0 & 0 & 0 & 0 & 0 & 0 & 0 & 0 & 0 & 0 \\
 0 & \mu _0 & 0 & 0 & 0 & 0 & 0 & 0 & 0 & 0 & 0 & 0 & 0 & 0 & 0 & 0 \\
 0 & 0 & \mu _0 & 0 & 0 & 0 & 0 & 0 & 0 & 0 & 0 & 0 & 0 & 0 & 0 & 0 \\
 0 & 0 & 0 & \mu _0+\mu _1 & 0 & 0 & 0 & 0 & 0 & 0 & 0 & 0 & 0 & 0 & 0 & 0 \\
 0 & 0 & 0 & 0 & \mu _0 & 0 & 0 & 0 & 0 & 0 & 0 & 0 & 0 & 0 & 0 & 0 \\
 0 & 0 & 0 & 0 & 0 & \mu _0+\mu _2 & 0 & 0 & 0 & 0 & 0 & 0 & 0 & 0 & 0 & 0 \\
 0 & 0 & 0 & 0 & 0 & 0 & \mu _0+\mu _4 & 0 & 0 & 0 & 0 & 0 & 0 & 0 & 0 & 0 \\
 0 & 0 & 0 & 0 & 0 & 0 & 0 & \mu _0 & 0 & 0 & 0 & 0 & 0 & 0 & 0 & 0 \\
 0 & 0 & 0 & 0 & 0 & 0 & 0 & 0 & \mu _0 & 0 & 0 & 0 & 0 & 0 & 0 & 0 \\
 0 & 0 & 0 & 0 & 0 & 0 & 0 & 0 & 0 & \mu _0+\mu _3 & 0 & 0 & 0 & 0 & 0 & 0 \\
 0 & 0 & 0 & 0 & 0 & 0 & 0 & 0 & 0 & 0 & \mu _0+\mu _5 & 0 & 0 & 0 & 0 & 0 \\
 0 & 0 & 0 & 0 & 0 & 0 & 0 & 0 & 0 & 0 & 0 & \mu _0 & 0 & 0 & 0 & 0 \\
 0 & 0 & 0 & 0 & 0 & 0 & 0 & 0 & 0 & 0 & 0 & 0 & \mu _0+\mu _6 & 0 & 0 & 0 \\
 0 & 0 & 0 & 0 & 0 & 0 & 0 & 0 & 0 & 0 & 0 & 0 & 0 & \mu _0 & 0 & 0 \\
 0 & 0 & 0 & 0 & 0 & 0 & 0 & 0 & 0 & 0 & 0 & 0 & 0 & 0 & \mu _0 & 0 \\
 0 & 0 & 0 & 0 & 0 & 0 & 0 & 0 & 0 & 0 & 0 & 0 & 0 & 0 & 0 & \mu _0 \\
\end{array}
\right). \nonumber 
       \end{eqnarray}

\end{enumerate}

\section{Conclusion}
\label{sec:conclusion}
In this work we have introduced and developed a technique to generate solutions to the $d$-simplex equations using Clifford algebras. This provides solutions to the  Yang-Baxter equation ($d=2$) and its higher dimensional counterpart, the tetrahedron equation ($d=3$). Their generalisations are far less studied. As far as we know there are few solutions for the $d$-simplex equations for $d\geq 4$ \cite{Hietarinta_1997,bardakov2022settheoretical}. The Clifford algebra method is unique as it solves all of these equations, including many of their variations. The set of solutions also form a linear space. The solutions vary with the representation of the local Hilbert space. We summarise this here.

Our method is based on two anticommuting operators $A$ and $B$ (generalised to a pair of anticommuting sets \eqref{eq:ABClifford}). These operators are shown to be realised by Clifford algebras of the form $\mathbf{CL}(p,q)$ (more generally $\mathbf{CL}(p,q,y)$ that includes nilpotent operators as well). The solutions are algebraic. They do not depend on the particular representation chosen for the Clifford algebras.
We find that for every pair of integers $(a,b)$, with either one of $a$ or $b$ being even, we can write down a solution of the $d=a+b$-simplex equation (See Sec. \ref{subsec:gend}). The space of solutions is linear. The solutions for the $d=2,3,4,5$ cases are summarised below.
\subsubsection*{Summary of the new solutions}
\label{subsubsec:summaryofsolutions}
\begin{enumerate}
    \item The simplest solution to the 2-simplex or the {\bf Yang-Baxter equation} is provided by the formula \eqref{eq:Rd2} (linear combination of types $(2,0)$ and $(0,2)$), while a more general solution can be found in the formula \eqref{eq:Rd2general}.
    \item For the 3-simplex or the {\bf tetrahedron equation}, the simplest product solution is given by the formula \eqref{eq:Rd3product} (type $(2,1)$).   It  can be slightly generalised to a non-product solution in \eqref{eq:Rd3} (linear combination of type $(2,1)$ operators). More general solutions are given by formulae \eqref{eq:Rd3general} and \eqref{eq:Rd3generalBA}.
    \item The {\bf 4-simplex equation} has two sets of product solutions given by \eqref{eq:Rd4product-1} (types $(4,0)$ and $(0,4)$) and \eqref{eq:Rd4product-2} (linear combination of types $(4,0)$ and $(0,4)$). A slightly more non-trivial solution, in \eqref{eq:R1d4}, paves way for a more general solution in \eqref{eq:R1d4general}. An analogous general solution to \eqref{eq:R2d4} (linear combination of type $(2,2)$ operators) can also be written down.
    \item The product solutions for the {\bf 5-simplex equation} are given by \eqref{eq:Rd5product-1} (type $(4,1)$) and \eqref{eq:Rd5product-2} (type $(3,2)$). The constituents of the latter are used to construct the more non-trivial solutions in \eqref{eq:R1d5} (linear combination of type $(4,1)$ operators) and \eqref{eq:R2d5} (linear combination of type $(3,2)$ operators) respectively. More general solutions can be written down following the logic used in the previous cases. 
    \item The most general solutions for each $d$ can be obtained by combining the different types. The prescription for this is given in Theorem 1 of Sec. \ref{subsec:gend}.
    \item As examples, the {\bf qubit representations} for the $d=2,3,4$ cases are shown in Sec. \ref{sec:qubitsolutions}. 
\end{enumerate}
The scope of the Clifford method is manifested in a series of Appendices. Their contents are summarised below.
\subsubsection*{Summary of content in the Appendices}
\label{subsubsec:summaryofappendices}
\begin{enumerate}
    \item For every even $d$ there are a number of $d$-simplex operators that correspond to the pair $(a,b)$, with both $a$ and $b$ odd. These operators solve a modified version of the $d$-simplex equations, where the right hand side of the equations is multiplied by an overall negative sign. We call these the {\bf anti-$d$-simplex equations}. Their solutions are the subject of Appendix \ref{app:antidsimplex}. The odd $d$ case is also discussed.
    \item The different forms of the {\bf tetrahedron equation} are solved using the Clifford algebra method in Appendix \ref{app:othertetrahedra}.
    \item Next we consider {\bf reflection equations} for $d=2,3$. These help model integrable systems on open manifolds. We show that these can be solved with the Clifford method in Appendix \ref{app:reflection}. We believe that the solutions to the higher simplex versions of the reflection equations can also be obtained in a similar manner.
    \item We show that there exists {\bf non-Clifford solutions} for the higher simplex equations in Appendix \ref{app:nonClifford}. With some restrictions the product Clifford solutions in the main text are special cases of these solutions.
    \item The Mathematica codes to verify the $d$-simplex solutions are shown in Appendix \ref{app:Mcodes}.
\end{enumerate}
We close with a few remarks on the potential applications of the higher simplex operators. There are three applications which are worth further study :
\begin{enumerate}
    \item An important application of the Yang-Baxter $R$-matrices is in finding novel integrable spin chains. The higher simplex operators give the framework for constructing the higher dimensional analogs. For example the two dimensional toric code of Kitaev can be obtained this way \cite{Khachatryan2015IntegrabilityIT}. The solutions constructed in this paper  provide a starting point to find these examples. These have implications for quantum phases in higher dimensions.
    \item In the recent decade there is a surge in interest for using the $R$-matrices as quantum gates in information processing. These are believed to result in less noisy quantum circuits due to the many integrals of motion. It would be interesting to use the higher simplex operators as quantum gates in quantum circuits. These can be achieved once we identify the unitary solutions among the higher simplex operators. This problem is far from obvious. Nevertheless we observe that there are already some unitary operators among the solutions studied in this work. The simplest solutions for each $d$, namely the operators that are of the factorised type, are all unitary as the generators of $\mathbf{CL}(p,q)$ are Hermitian and they square to $\mathbb{1}$. These include operators of the type \eqref{eq:Rd3product}, \eqref{eq:Rd4product-1}, \eqref{eq:Rd4product-2}, \eqref{eq:Rd5product-1}, \eqref{eq:Rd5product-2} and their higher simplex analogs. A more non-trivial solution for the $d=4$ case is given by 
    \begin{eqnarray}
        R_{ijkl} = \frac{1}{2}\left[A_iB_jA_kB_l + B_iA_jA_kB_l -  A_iB_jB_kA_l + B_iA_jB_kA_l \right].
    \end{eqnarray}
    \item Two applications in mathematics : The higher simplex operators can help find the invariant polynomials corresponding to higher dimensional versions of knots. As the higher simplex equations appear as constraint equations in higher categories, the solutions shown in this work can help construct examples of such structures.
\end{enumerate}

\section*{Acknowledgments}
We are grateful to Kun Zhang, Vivek Kumar Singh and Akash Sinha for discussions. We also thank Jarmo Hietarinta, Atsuo Kuniba, Jacques H. H. Perk and Junya Yagi for comments and suggestions.
This research is funded by the U.S. Department of Energy, Office of Science, National Quantum Information Science Research Centers, Co-Design Center for Quantum Advantage under Contract No. DE-SC0012704.

\appendix

\section{Proof that $\alpha=\pm 1$ to satisfy the $d$-simplex equation }
\label{app:ABalphad3}
Assume that \eqref{eq:ABalpha} holds. We will show that $\alpha$ has to be $\pm 1$ for the $d$-simplex equations to hold. For the following proofs we will use the obvious adaptation of the notation introduced in \ref{subsec:d3}. We begin with the 3-simplex operator in \eqref{eq:Rd3}. 
\begin{eqnarray}
   & & R_{123}R_{145}R_{246}R_{356} \nonumber \\
   & = & A_3A_5B_6\left(\frac{1}{\alpha},+,+\right)_{123}\left(\frac{1}{\alpha},+,+\right)_{145}\left(+,\alpha,\alpha\right)_{246} \nonumber \\
   & + & A_3B_5A_6\left(\frac{1}{\alpha},+,+\right)_{123}\left(+,\alpha,\alpha\right)_{145}\left(\frac{1}{\alpha},+,+\right)_{246} \nonumber \\
   & + & B_3A_5A_6\left(+,\alpha,\alpha\right)_{123}\left(\frac{1}{\alpha},+,+\right)_{145}\left(\frac{1}{\alpha},+,+\right)_{246} \nonumber \\  
   & = & \frac{1}{\alpha^2}R_{356}\left(+,\alpha,\alpha\right)_{123}\left(+,\alpha,\alpha\right)_{145}\left(+,\alpha,\alpha\right)_{246} \nonumber \\
   & = & \frac{1}{\alpha^2}R_{356} \left[A_2A_4B_6\left(+,+,\alpha\right)_{123}\left(+,+,\alpha\right)_{145} \right. \nonumber \\
   & + & \left. \alpha A_2B_4A_6\left(+,+,\alpha\right)_{123}\left(\alpha,\alpha,\alpha^2\right)_{145}  \right. \nonumber \\
   & + & \left. \alpha B_2A_4A_6\left(\alpha,\alpha,\alpha^2\right)_{123}\left(+,+,\alpha\right)_{145} \right] \nonumber \\
   & = & \frac{1}{\alpha^2}R_{356}\left(+,\alpha^2,\alpha^2\right)_{246}\left(+,+,\alpha\right)_{123}\left(+,+,\alpha\right)_{145} \nonumber \\
   & = & \frac{1}{\alpha^2}R_{356}\left(+,\alpha^2,\alpha^2\right)_{246}\left[A_1A_4B_5\left(+,+,+\right)_{123} \right. \nonumber \\
   & + & \left. A_1B_4A_5\left(+,+,+\right)_{123}  \right. \nonumber \\
   & + & \left. \alpha B_1B_4A_5\left(\alpha,\alpha,\alpha\right)_{123} \right] \nonumber \\
   & = & \frac{1}{\alpha^2}R_{356}\left(+,\alpha^2,\alpha^2\right)_{246}\left(+,+,\alpha^2\right)_{145}R_{123} \nonumber \\
   & = & R_{356}R_{246}R_{145}R_{123}~;~\textrm{iff}~\alpha=\pm 1.
\end{eqnarray}
Next we prove the same for the $(4,0)$ type 4-simplex operator in \eqref{eq:R1d4}.
\begin{eqnarray}
   & & R_{1234}R_{1567}R_{2589}R_{368,10}R_{479,10} \nonumber \\
   & = & \left(+, \alpha^4\right)_{479,10}\left( +, \frac{1}{\alpha}\right)_{1234}\left( +, \frac{1}{\alpha}\right)_{1567}\left( +, \frac{1}{\alpha}\right)_{2589}\left( +, \frac{1}{\alpha}\right)_{368,10} \nonumber \\
   & = & \left(+, \alpha^4\right)_{479,10}\left(+, \alpha^2\right)_{368,10}\left( +, \frac{1}{\alpha^2}\right)_{1234}\left( +, \frac{1}{\alpha^2}\right)_{1567}\left( +, \frac{1}{\alpha^2}\right)_{2589} \nonumber \\
   & = & \left(+, \alpha^4\right)_{479,10}\left(+, \alpha^2\right)_{368,10}R_{2589}\left( +, \frac{1}{\alpha^3}\right)_{1234}\left( +, \frac{1}{\alpha^3}\right)_{1567} \nonumber \\
   & = & \left(+, \alpha^4\right)_{479,10}\left(+, \alpha^2\right)_{368,10}R_{2589}\left( +, \frac{1}{\alpha^2}\right)_{1567}\left( +, \frac{1}{\alpha^4}\right)_{1234} \nonumber \\
   & = & R_{479,10}R_{368,10}R_{2589}R_{1567}R_{1234}~;~\textrm{iff}~\alpha=\pm 1.
\end{eqnarray}

We now turn to the $(2,2)$ type 4-simplex operator in \eqref{eq:R2d4}.
\begin{eqnarray}
    & & R_{1234}R_{1567}R_{2589}R_{368,10}R_{479,10} \nonumber \\
    & = & \frac{1}{\alpha^2}R_{479,10}\left(+,+,+,\alpha,\alpha,\alpha \right)_{1234}\left(+,+,+,\alpha,\alpha,\alpha \right)_{1567} \nonumber \\ & \times & \left(+,+,+,\alpha,\alpha,\alpha \right)_{2589}\left(+,+,+,\alpha,\alpha,\alpha \right)_{368,10} \nonumber \\
    & = & \frac{1}{\alpha^2}R_{479,10}\left(\frac{1}{\alpha^2},\frac{1}{\alpha^2},\frac{1}{\alpha^2},+,+,+ \right)_{368,10}\left(+,\alpha,\alpha,\alpha,\alpha,\alpha^2 \right)_{1234} \nonumber \\ & \times & \left(+,\alpha,\alpha,\alpha,\alpha,\alpha^2 \right)_{1567}\left(+,\alpha,\alpha,\alpha,\alpha,\alpha^2 \right)_{2589} \nonumber \\
     & = & \frac{1}{\alpha^2}R_{479,10}\left(\frac{1}{\alpha^2},\frac{1}{\alpha^2},\frac{1}{\alpha^2},+,+,+ \right)_{368,10}\left(+,\alpha^2,\alpha^2,\alpha^2,\alpha^2,\alpha^4 \right)_{2589} \nonumber \\ & \times & \left(+,+,\alpha,+,\alpha,\alpha \right)_{1234}\left(+,+,\alpha,+,\alpha,\alpha\right)_{1567} \nonumber \\
      & = & \frac{1}{\alpha^2}R_{479,10}\left(\frac{1}{\alpha^2},\frac{1}{\alpha^2},\frac{1}{\alpha^2},+,+,+ \right)_{368,10}\left(+,\alpha^2,\alpha^2,\alpha^2,\alpha^2,\alpha^4 \right)_{2589} \nonumber \\ & \times & \left(+,+,\alpha^2,+,\alpha^2,\alpha^2\right)_{1567}R_{1234} \nonumber \\
      & = & R_{479,10}R_{368,10}R_{2589}R_{1567}R_{1234}~;~\textrm{iff}~\alpha=\pm 1.
\end{eqnarray}
The above proofs go through for the more general 3- and 4-simplex operators in \eqref{eq:Rd3general}, \eqref{eq:R1d4general} as well. Similar proofs can be worked out for higher $d$, which we leave to the interested reader.

\section{The anti-$d$-simplex equations and its solutions}
\label{app:antidsimplex}
The $d$-simplex operators corresponding to the pairs $(a,b)$ for odd $a$ and $b$ satisfy the anti-$d$-simplex equations. These are well defined for even $d$. In this appendix we will define these equations and prove that these pairs solve them.

The anti-$d$-simplex equations are essentially the same as the $d$-simplex equations except for an overall negative sign on the right hand sides of the latter. This is the case when $d$ is even. The situation changes when $d$ is odd, which will be discussed separately. 

We begin with the even case. The anti-$2$-simplex equation or the anti-Yang-Baxter equation is given by 
\begin{equation}\label{eq:antid2simplex}
    R_{12}R_{13}R_{23} = -R_{23}R_{13}R_{12}.
\end{equation}
Note the minus sign in the right hand side. This equation  is solved by $A_iB_j$, $B_iA_j$ and their linear combination\footnote{When $A^2=B^2=\mathbb{1}$, it turns out that this operator is non-invertible and satisfies $$ R_{12}R_{13}R_{23}=0.$$ A better ansatz, that is also invertible, is given by \begin{equation}
    R_{ij} = A_iB_j + \mu~B_iA_j.
\end{equation} The proof that it satisfies the anti-Yang-Baxter equation is the same as the one shown above. },
\begin{equation}\label{eq:antiRd2}
    R_{ij} = A_iB_j + B_iA_j.
\end{equation}
The proof, using the shorthand notation of Sec. \ref{subsec:d3}, goes as 
\begin{eqnarray}
   R_{12}R_{13}R_{23} & = & R_{23}\left(-,+\right)_{12}\left(+,-\right)_{13} \nonumber \\
   & = & R_{23}R_{13}\left(-,-\right)_{12} \nonumber \\
   & = & -R_{23}R_{13}R_{12}.
\end{eqnarray}
The anti-4-simplex equation is given by 
\begin{eqnarray}
     R_{1234}R_{1567}R_{2589}R_{368,10}R_{479,10} 
     =  -R_{479,10}R_{368,10}R_{2589}R_{1567}R_{1234}. \label{eq:antid4simplex}
\end{eqnarray}
This is solved by
\begin{equation}\label{eq:antiRd4}
    R_{ijkl} = A_iA_jA_kB_l + A_iA_jB_kA_l + A_iB_jA_kA_l + B_iA_jA_kA_l,
\end{equation}
corresponding to the pair $(3,1)$. The pair $(1,3)$ gives another solution with the $A$ and $B$ operators interchanged in \eqref{eq:antiRd4}.
Using the shorthand notations introduced in Sec. \ref{subsec:d3} the proof goes as 
\begin{eqnarray}
  & &  R_{1234}R_{1567}R_{2589}R_{368,10}R_{479,10} \nonumber \\
  & = & R_{479,10}\left(-,+,+,+\right)_{1234}\left(-,+,+,+\right)_{1567}\left(-,+,+,+\right)_{2589}\left(+,-,-,-\right)_{368,10} \nonumber \\
  & = & R_{479,10}R_{368,10}\left(-,-,+,+\right)_{1234}\left(-,-,+,+\right)_{1567}\left(-,-,+,+\right)_{2589} \nonumber \\
  & = & R_{479,10}R_{368,10}R_{2589}\left(-,-,-,+\right)_{1234}\left(+,+,+,-\right)_{1567} \nonumber \\
  & = & R_{479,10}R_{368,10}R_{2589}R_{1567}\left(-,-,-,-\right)_{1234} \nonumber \\
  & = & -R_{479,10}R_{368,10}R_{2589}R_{1567}R_{1234}.
\end{eqnarray}
The extensions to other even $d$ is straightforward. The linear structure in the space of solutions holds for these equations as well.

When $d$ is odd the $d$-simplex operator corresponds to the pair $(d,0)$. For example when $d=3$ this corresponds to 
\begin{equation}\label{eq:antiRd3}
    R_{ijk} = A_iA_jA_k + B_iB_jB_k. 
\end{equation}
This solves the anti-3-simplex equation or the anti-tetrahedron equation
\begin{equation}\label{eq:antid3simplex}
    R_{123}R_{145}R_{246}R_{356} = R^{(-)}_{356}R^{(-)}_{246}R^{(-)}_{145}R^{(-)}_{123},
\end{equation}
where 
\begin{equation}
    R^{(-)}_{ijk} = A_iA_jA_k - B_iB_jB_k. 
\end{equation}
The proof is as follows :
\begin{eqnarray}
    R_{123}R_{145}R_{246}R_{356} & = & R^{(-)}_{356}\left(+,-\right)_{123}\left(+,-\right)_{145}\left(+,-\right)_{246} \nonumber \\
    & = & R^{(-)}_{356}R^{(-)}_{246}\left(+,+\right)_{123}\left(+,+\right)_{145} \nonumber \\
    & = & R^{(-)}_{356}R^{(-)}_{246}R^{(-)}_{145}\left(+,-\right)_{123} \nonumber \\
    & = & R^{(-)}_{356}R^{(-)}_{246}R^{(-)}_{145}R^{(-)}_{123}.
\end{eqnarray}
The extension to larger odd $d$ values is obvious. It is worth noting that the anti-$d$-simplex equation, for odd $d$, is more natural when a spectral parameter is included into the $d$-simplex operator \eqref{eq:antiRd3},
\begin{equation}\label{eq:antiRd3SP}
    R_{ijk} = A_iA_jA_k + \mu_{ijk}~B_iB_jB_k. 
\end{equation}
This satisfies a spectral parameter dependent version of \eqref{eq:antid3simplex}
\begin{eqnarray}\label{eq:antid3simplexSP}
& & R_{123}\left(\mu_{123}\right)R_{145}\left(\mu_{145}\right)R_{246}\left(\mu_{246}\right)R_{356}\left(\mu_{356}\right)\nonumber \\
& = & R_{356}\left(-\mu_{356}\right)R_{246}\left(-\mu_{246}\right)R_{145}\left(-\mu_{145}\right)R_{123}\left(-\mu_{123}\right).
\end{eqnarray}
This easily generalises to higher values of odd $d$.

\section{Solutions for other forms of the tetrahedron equation}
\label{app:othertetrahedra}
In this section we will show that ansatzes constructed using Clifford algebras solve the edge form of the tetrahedron equation \eqref{eq:tetrahedronEdgeForm} and the quantized Yang-Baxter equation \eqref{eq:tetrahedronYBconjugator}. For both cases we will use the ansatz in \eqref{eq:Rd3} which we repeat here :
\begin{equation}
    R_{ijk} = A_iA_jB_k + A_iB_jA_k + B_iA_jA_k.
\end{equation}

\subsection*{Proof that \eqref{eq:Rd3} solves \eqref{eq:tetrahedronEdgeForm}}
We invoke the shorthand notation introduced in Sec. \ref{subsec:d3} for this case.
\begin{eqnarray}
  & &  R_{123}R_{124}R_{134}R_{234} \nonumber \\
  & = & R_{234}\left(-,-,+\right)_{123}\left(+,+,-\right)_{124}\left(+,+,-\right)_{134} \nonumber \\
  & = & R_{234}\left[A_1A_3B_4\left(+,-,-\right)_{123}\left(+,-,-\right)_{124} \right. \nonumber \\
  & + & \left. A_1B_3A_4\left(-,+,+\right)_{123}\left(-,+,+\right)_{124} \right. \nonumber \\
  & - & \left. B_1A_3A_4\left(-,+,+\right)_{123}\left(+,-,-\right)_{124} \right] \nonumber \\
  & = & R_{234}R_{134} \left(+,-,-\right)_{123}\left(+,-,-\right)_{124} \nonumber \\
  & = & R_{234}R_{134} \left[ A_1A_2B_4\left(+,+,+\right)_{123} \right. \nonumber \\
  & - & \left. A_1B_2A_4\left(-,-,-\right)_{123} \right. \nonumber \\
  & - & \left. B_1A_2A_4\left(-,-,-\right)_{123} \right] \nonumber \\
  & = & R_{234}R_{134}R_{124}R_{123}.
\end{eqnarray}
\subsection*{Proof that \eqref{eq:Rd3} solves \eqref{eq:tetrahedronYBconjugator}}
Using the shorthand notation in Sec. \ref{subsec:d3} the proof goes as :
\begin{eqnarray}
   & &  R_{124}R_{135}R_{236}R_{456} \nonumber \\
   & = & R_{456}\left(-,+,+\right)_{124}\left(-,+,+\right)_{135}\left(+,-,-\right)_{236} \nonumber \\
   & = & R_{456}\left[A_2A_3B_6 \left(-,-,+\right)_{124}\left(-,-,+\right)_{135} \right. \nonumber \\
   & - & \left.A_2B_3A_6 \left(-,-,+\right)_{124}\left(+,+,-\right)_{135}  \right. \nonumber \\
   & - & \left.B_2A_3A_6 \left(+,+,-\right)_{124}\left(-,-,+\right)_{135}  \right] \nonumber \\
   & = & R_{456}R_{236}\left(-,-,+\right)_{124}\left(-,-,+\right)_{135} \nonumber \\
   & = & R_{456}R_{236}\left[-A_1A_3B_5 \left(-,-,-\right)_{124} \right. \nonumber \\
   & - & \left. A_1B_3A_5 \left(-,-,-\right)_{124}\right. \nonumber \\
   & + & \left. B_1A_3A_5 \left(+,+,+\right)_{124}\right] \nonumber \\
   & = & R_{456}R_{236}R_{135}R_{124}.
\end{eqnarray}

\section{Solutions of reflection equations}
\label{app:reflection}
We now show that the operators generated from Clifford algebras can also solve the reflection equations \cite{kuniba2022quantum}. This is shown for the $d=2$ and $d=3$ cases.

\subsection*{$d=2$} 
The constant form of the reflection equation in the $d=2$ case is given by
\begin{equation}\label{eq:d2reflection}
    R_{12}K_2R_{21}K_1 = K_1R_{12}K_2R_{21}.
\end{equation}
This is solved by the ansatz
\begin{eqnarray}\label{eq:Rd2Kd2}
    R_{ij} = A_iA_j + B_iB_j~;~K_j=A_j+B_j,
\end{eqnarray}
with anticommuting $A$ and $B$.
Adapting the shorthand notation of Sec. \ref{subsec:d3} this can be proved as follows :
\begin{eqnarray}
    R_{12}K_2R_{21}K_1 & = & K_1\left(+,-\right)_{12}\left(+,+\right)_{2}\left(+,-\right)_{21} \nonumber \\
    & = & K_1R_{12}\left(+,-\right)_{21}\left(+,-\right)_{2} \nonumber \\
    & = & K_1R_{12}K_2\left(+,+\right)_{21} \nonumber \\
    & = & K_1R_{12}K_2R_{21}.
\end{eqnarray}
The ansatz in \eqref{eq:Rd2Kd2} can be generalised by including arbitrary linear combinations and more number of operators.

\subsection*{$d=3$}
The constant form of the $d=3$ reflection equation is given by 
\begin{eqnarray}\label{eq:d3reflection}
    & & R_{689}K_{3579}R_{249}R_{258}K_{1478}K_{1236}R_{456} \nonumber \\
    & = & R_{456}K_{1236}K_{1478}R_{258}R_{249}K_{3579}R_{689}.
\end{eqnarray}
This equation is solved by the ansatz
\begin{eqnarray}\label{eq:Rd3Kd3}
    R_{ijk} & = & A_iA_jB_k + A_iB_jA_k + B_iA_jA_k \nonumber \\
    K_{ijkl} & = & A_iA_jA_kA_l + B_iB_jB_kB_l.
\end{eqnarray}
Using the shorthand notation of Sec. \ref{subsec:d3} we can prove this as follows :
\begin{eqnarray}
     & & R_{689}K_{3579}R_{249}R_{258}K_{1478}K_{1236}R_{456} \nonumber \\
     & = & R_{456}\left(-,-,+\right)_{689}\left(+,-\right)_{3579}\left(+,-,+\right)_{249}\left(+,-,+\right)_{258}\left(-,+\right)_{1478}\left(-,+\right)_{1236} \nonumber \\
     & = & R_{456}K_{1236}\left(+,+,+\right)_{689}\left(+,+\right)_{3579}\left(+,-,-\right)_{249}\left(+,-,-\right)_{258}\left(+,+\right)_{1478} \nonumber \\
     & = & R_{456}K_{1236}K_{1478}\left(+,-,+\right)_{689}\left(+,-\right)_{3579}\left(+,+,-\right)_{249}\left(-,-,-\right)_{258} \nonumber \\
     & = & R_{456}K_{1236}K_{1478}R_{258}\left(+,+,+\right)_{689}\left(+,+\right)_{3579}\left(+,+,+\right)_{249} \nonumber \\
     & = & R_{456}K_{1236}K_{1478}R_{258}R_{249}\left(+,-,-\right)_{689}\left(-,+\right)_{3579} \nonumber \\
     & = & R_{456}K_{1236}K_{1478}R_{258}R_{249}K_{3579}R_{689}.
\end{eqnarray}
The ansatz in \eqref{eq:Rd3Kd3} generalises with the inclusion of arbitrary linear combinations and more number of operators from the sets in \eqref{eq:ABClifford}.

\section{Non-Clifford solutions}
\label{app:nonClifford}
Now we have a brief look at the possibility for a non-Clifford solution to the $d$-simplex equation. We consider the $d=2$, $d=3$ and $d=4$ cases before we generalise to an arbitrary $d$. 
\paragraph{\bf{$d=2$} :} Consider the ansatz
\begin{equation}\label{eq:Rd2nonClifford}
    R_{ij} = A_iB_j,
\end{equation}
with the operators $A$ and $B$ satisfying the relation 
\begin{equation}\label{eq:ABnonClifford}
    AB = \alpha BA~;~\alpha\in\mathbb{C}.
\end{equation}
This satisfies the 2-simplex or the Yang-Baxter equation when $\alpha=1$. This case is included in \cite{padmanabhan2024integrability}.
\paragraph{\bf{$d=3$} :} Now we take the ansatz as
\begin{equation}\label{eq:Rd3nonClifford}
    R_{ijk} = A_iB_jC_k,
\end{equation}
with the operators $A$, $B$ and $C$ satisfying
\begin{equation}\label{eq:ABCnonClifford}
    AB = \alpha BA~;~AC = \beta CA~;~BC = \gamma CB~;~ \alpha,\beta,\gamma\in\mathbb{C}.
\end{equation}
This satisfies the 3-simplex or the tetrahedron equation when 
\begin{equation}\label{eq:alphabetagamma}
    \alpha\beta\gamma = 1.
\end{equation}
A non-trivial choice for the parameters are the third roots of unity.
\paragraph{\bf{$d=4$} :} This solution will help us identify the pattern for an arbitrary $d$. We introduce new notation to make things simpler. This is seen in the ansatz
\begin{equation}\label{eq:Rd4nonClifford}
    R_{ijkl} =  \left(A_1\right)_i\left(A_2\right)_j\left(A_3\right)_k\left(A_4\right)_l.
\end{equation}
Here the indices $i$, $j$, $k$ and $l$ denote the local Hilbert spaces appearing in the $d$-simplex operators. The indices 1,2,3, and 4 index the operators of the algebra. We will assume that no confusion will arise to the reader due to this. The algebra satisfied by the $A_p$ operators is
\begin{equation}\label{eq:ApnonClifford}
    A_pA_q = \alpha_{pq} A_qA_p~;~p<q\in\{1,2,3,4\}.
\end{equation}
After some simple computations we can verify that \eqref{eq:Rd4nonClifford} satisfies the 4-simplex equation \eqref{eq:d4simplex} when 
\begin{equation}
    \mathop{\prod\limits_{p,q=1}^4}_{p<q}~\alpha_{pq} = 1.
\end{equation}
A non-trivial solution is presented by the sixth roots of unity. 
This notation helps us to generalise to an arbitrary $d$. 
\paragraph{\bf{General $d$} :} The ansatz for the $d$-simplex operator is given by 
\begin{equation}\label{eq:RdgeneralnonClifford}
    R_{i_1\cdots i_d} = \bigotimes_{p=1}^d\left(A_p\right)_{i_p},
\end{equation}
with the $A_p$ operators satisfying
\begin{equation}\label{eq:ApnonCliffordgenerald}
    A_pA_q = \alpha_{pq} A_qA_p~;~p<q\in\{1,\cdots, d\}.
\end{equation}
This satisfies the $d$-simplex equation when
\begin{equation}
    \mathop{\prod\limits_{p,q=1}^d}_{p<q}~\alpha_{pq} = 1.
\end{equation}
A non-trivial solution is given by the $\binom{d}{2}$th roots of unity.

\section{Mathematica Codes}
\label{app:Mcodes}
We consider the first two cases of Sec. \ref{subsec:clifford} in what follows. This Appendix should be read along with the qubit solutions presented in Sec. \ref{sec:qubitsolutions}.

\paragraph{{\bf Case 1} -}
{\it {\bf 3-simplex operators} :}
Below we show the Mathematica code to verify that the 3-simplex operator in \eqref{eq:Rd3matrix} satisfies the spectral parameter dependent 3-simplex equation \eqref{eq:d3simplexSP}. The first(second) box is the input(output). The latter includes the choice of the randomly generated 15 parameters ($\mu$) taken randomly from integers between -10 and 10. It also includes the Boolean value of the veracity of the 3-simplex equation. The $R$-matrices contain 6 parameters, the first three ($\alpha$, $\beta$, $\gamma$) specify the indices on which it acts non-trivially and the remaining are chosen from the 15 $\mu$ parameters. Symbolic computations also confirms that the operator in \eqref{eq:Rd3matrix} is a solution.

\begin{tcolorbox}
\begin{doublespace}
\noindent\(\pmb{\text{sigmaX}=\text{SparseArray}@\text{PauliMatrix}[1];}\\
\pmb{\text{sigmaZ}=\text{SparseArray}@\text{PauliMatrix}[3];}\\
\pmb{\mu=\text{Table}[\text{RandomInteger}[\{-10,10\}],\{i,1,15\}]}\\
\pmb{\text{R3d}[\alpha \_,\beta \_,\gamma \_,\text{i$\_$},\text{j$\_$},\text{k$\_$}]\text{:=}} \\
\pmb{\mu[[i]]*\text{KroneckerProduct}\text{@@}\text{ReplacePart}[\text{ConstantArray}[\text{IdentityMatrix}[2],6],} \\
\pmb{\{\alpha
\to \text{sigmaX},\beta \to \text{sigmaX},\gamma \to \text{sigmaZ}\}]+}\\
\pmb{\mu[[j]]*\text{KroneckerProduct}\text{@@}\text{ReplacePart}[\text{ConstantArray}[\text{IdentityMatrix}[2],6],} \\ 
\pmb{\{\alpha \to \text{sigmaX},\beta
\to \text{sigmaZ},\gamma \to \text{sigmaX}\}] + }\\
\pmb{\mu[[k]]*\text{KroneckerProduct}\text{@@}\text{ReplacePart}[\text{ConstantArray}[\text{IdentityMatrix}[2],6],} \\ 
\pmb{\{\alpha \to \text{sigmaZ},\beta
\to \text{sigmaX},\gamma \to \text{sigmaX}\}];}\\
\pmb{\text{isEqual}=\text{Simplify}[\text{R3d}[1,2,3,8,11,12].\text{R3d}[1,4,5,3,7,9].} \\ 
\pmb{\text{R3d}[2,4,6,13,10,8].\text{R3d}[3,5,6,14,10,12]\text{==}}\\
\pmb{\text{R3d}[3,5,6,14,10,12].\text{R3d}[2,4,6,13,10,8].\text{R3d}[1,4,5,3,7,9].\text{R3d}[1,2,3,8,11,12]];}\\
\pmb{\text{(*Print result*)}}\\
\pmb{\text{If}[\text{isEqual},\text{Print}[\text{{``}True{''}}],\text{Print}[\text{{``}False{''}}]]}\)
\end{doublespace}

\end{tcolorbox}

\begin{tcolorbox}
  \begin{doublespace}
\noindent\(\{7,2,6,4,8,-8,7,10,-7,-10,-3,6,8,9,-2\}\)
\end{doublespace}

\noindent\(\text{True}\)
\end{tcolorbox}

{\it {\bf 4-simplex operators} :}
The boxes below show the Mathematica code for the verification that a linear combination of the 4-simplex operators of the $(4,0)$ and $(2,2)$  types satisfy the spectral parameter dependent 4-simplex equation of \eqref{eq:d4simplexSPgeneral}. In the code each of the five 4-simplex operators on each side of this equation depend on 12 parameters. The first four parameters ($\alpha$, $\beta$, $\gamma$, $\delta$) specify the position of the $X$ and $Z$ operators in the tensor product. The remaining 8 parameters are the coefficients appearing in the linear combination. These are picked randomly from integers between -10 and 10. The output shows these parameters and the Boolean value for the truth of the 4-simplex equation. The coefficients can take real or complex values as well. This requires different precision levels.

\begin{tcolorbox}
   \begin{doublespace}
\noindent\(\pmb{\text{sigmaX}=\text{SparseArray}@\text{PauliMatrix}[1];}\\
\pmb{\text{sigmaZ}=\text{SparseArray}@\text{PauliMatrix}[3];}\\
\pmb{\mu=\text{Table}[\text{RandomInteger}[\{-10,10\}],\{i,1,15\}]}\\
\pmb{\text{R4d}[\alpha \_,\beta \_,\gamma \_,\delta \_,\text{i$\_$},\text{j$\_$},\text{k$\_$},\text{l$\_$},\text{m$\_$},\text{n$\_$},\text{p$\_$},\text{q$\_$}]\text{:=}}\\
\pmb{\mu[[i]]*\text{KroneckerProduct}\text{@@}\text{ReplacePart}[\text{ConstantArray}[\text{IdentityMatrix}[2],10],} \\ 
\pmb{\{\alpha \to \text{sigmaX},\beta
\to \text{sigmaX},\gamma \to \text{sigmaZ},\delta \to \text{sigmaZ}\}]+}\\
\pmb{\mu[[j]]*\text{KroneckerProduct}\text{@@}\text{ReplacePart}[\text{ConstantArray}[\text{IdentityMatrix}[2],10],} \\ 
\pmb{\{\alpha \to \text{sigmaX},\beta
\to \text{sigmaZ},\gamma \to \text{sigmaX},\delta \to \text{sigmaZ}\}]+}\\
\pmb{\mu[[k]]*\text{KroneckerProduct}\text{@@}\text{ReplacePart}[\text{ConstantArray}[\text{IdentityMatrix}[2],10],} \\ 
\pmb{\{\alpha \to \text{sigmaZ},\beta
\to \text{sigmaX},\gamma \to \text{sigmaX},\delta \to \text{sigmaZ}\}]+}\\
\pmb{\mu[[l]]*\text{KroneckerProduct}\text{@@}\text{ReplacePart}[\text{ConstantArray}[\text{IdentityMatrix}[2],10],} \\ 
\pmb{\{\alpha \to \text{sigmaX},\beta
\to \text{sigmaZ},\gamma \to \text{sigmaZ},\delta \to \text{sigmaX}\}] +}\\
\pmb{\mu[[m]]*\text{KroneckerProduct}\text{@@}\text{ReplacePart}[\text{ConstantArray}[\text{IdentityMatrix}[2],10],} \\ 
\pmb{\{\alpha \to \text{sigmaZ},\beta
\to \text{sigmaX},\gamma \to \text{sigmaZ},\delta \to \text{sigmaX}\}] +}\\
\pmb{\mu[[n]]*\text{KroneckerProduct}\text{@@}\text{ReplacePart}[\text{ConstantArray}[\text{IdentityMatrix}[2],10],} \\ 
\pmb{\{\alpha \to \text{sigmaZ},\beta
\to \text{sigmaZ},\gamma \to \text{sigmaX},\delta \to \text{sigmaX}\}] +}\\
\pmb{\mu[[p]]*\text{KroneckerProduct}\text{@@}\text{ReplacePart}[\text{ConstantArray}[\text{IdentityMatrix}[2],10],} \\ 
\pmb{\{\alpha \to \text{sigmaX},\beta
\to \text{sigmaX},\gamma \to \text{sigmaX},\delta \to \text{sigmaX}\}]+}\\
\pmb{\mu[[q]]*\text{KroneckerProduct}\text{@@}\text{ReplacePart}[\text{ConstantArray}[\text{IdentityMatrix}[2],10],} \\ 
\pmb{\{\alpha \to \text{sigmaZ},\beta
\to \text{sigmaZ},\gamma \to \text{sigmaZ},\delta \to \text{sigmaZ}\}];}\\
\pmb{\text{Simplify}[\text{R4d}[1,2,3,4,11,12,8,7,9,10,5,13].\text{R4d}[1,5,6,7,11,15,2,5,6,7,8,14].} \\ 
\pmb{\text{R4d}[2,5,8,9,12,15,8,13,11,15,12,10].\text{R4d}[3,6,8,10,13,6,1,4,14,9,5,7].} \\ 
\pmb{\text{R4d}[4,7,9,10,14,7,2,5,8,10,11,13]==}\\
\pmb{\text{R4d}[4,7,9,10,14,7,2,5,8,10,11,13].\text{R4d}[3,6,8,10,13,6,1,4,14,9,5,7].} \\ 
\pmb{\text{R4d}[2,5,8,9,12,15,8,13,11,15,12,10].\text{R4d}[1,5,6,7,11,15,2,5,6,7,8,14].} \\ 
\pmb{\text{R4d}[1,2,3,4,11,12,8,7,9,10,5,13]]}\)
\end{doublespace}

\end{tcolorbox}

\begin{tcolorbox}
    \begin{doublespace}
\noindent\(\{9,-3,7,-8,8,2,6,-7,-6,0,7,-2,7,-10,-7\}\)

\noindent\(\text{True}\)
\end{doublespace}
\end{tcolorbox}

{\it {\bf 5-simplex operators} :} As an example we provide the Mathematica code for the $(4,1)$ type 5-simplex operator :

\begin{tcolorbox}
    \begin{align*}
&\pmb{\text{sigmaX}=\text{SparseArray}@\text{PauliMatrix}[1];}\\
&\pmb{\text{sigmaZ}=\text{SparseArray}@\text{PauliMatrix}[3];}\\
&\pmb{\mu=\text{Table}[\text{RandomInteger}[\{-10,10\}],\{i,1,15\}]}\\
&\pmb{\text{R5d1}[\alpha \_,\beta \_,\gamma \_,\delta \_,\eta \_,\text{i$\_$},\text{j$\_$},\text{k$\_$},\text{l$\_$},\text{m$\_$}]\text{:=}}\\
&\pmb{\mu[[i]]*\text{KroneckerProduct}\text{@@}\text{ReplacePart}[\text{ConstantArray}[\text{IdentityMatrix}[2],15],} \\ 
&\pmb{\{\alpha \to \text{sigmaX},\beta
\to \text{sigmaX},\gamma \to \text{sigmaX},\delta \to \text{sigmaX},\eta \to \text{sigmaZ}\}]+}\\
&\pmb{\mu[[j]]*\text{KroneckerProduct}\text{@@}\text{ReplacePart}[\text{ConstantArray}[\text{IdentityMatrix}[2],15],} \\ 
&\pmb{\{\alpha \to \text{sigmaX},\beta
\to \text{sigmaX},\gamma \to \text{sigmaX},\delta \to \text{sigmaZ},\eta \to \text{sigmaX}\}]+}\\
&\pmb{\mu[[k]]*\text{KroneckerProduct}\text{@@}\text{ReplacePart}[\text{ConstantArray}[\text{IdentityMatrix}[2],15],} \\ 
&\pmb{\{\alpha \to \text{sigmaX},\beta
\to \text{sigmaX},\gamma \to \text{sigmaZ},\delta \to \text{sigmaX},\eta \to \text{sigmaX}\}]+}\\
&\pmb{\mu[[l]]*\text{KroneckerProduct}\text{@@}\text{ReplacePart}[\text{ConstantArray}[\text{IdentityMatrix}[2],15],} \\ 
&\pmb{\{\alpha \to \text{sigmaX},\beta
\to \text{sigmaZ},\gamma \to \text{sigmaX},\delta \to \text{sigmaX},\eta \to \text{sigmaX}\}] +}\\
&\pmb{\mu[[m]]*\text{KroneckerProduct}\text{@@}\text{ReplacePart}[\text{ConstantArray}[\text{IdentityMatrix}[2],15],} \\ 
&\pmb{\{\alpha \to \text{sigmaZ},\beta
\to \text{sigmaX},\gamma \to \text{sigmaX},\delta \to \text{sigmaX},\eta \to \text{sigmaX}\}];}\\
&\pmb{\text{Simplify}[\text{R5d1}[1,2,3,4,5,7,9,10,15,12].\text{R5d1}[1,6,7,8,9,13,14,12,1,2].} \\ 
&\pmb{\text{R5d1}[2,6,10,11,12,8,9,10,15,12].\text{R5d1}[3,7,10,13,14,6,4,11,14,13].} \\ 
&\pmb{\text{R5d1}[4,8,11,13,15,2,3,9,10,11].\text{R5d1}[5,9,12,14,15,3,6,14,11,1]\text{==}}\\
&\pmb{\text{R5d1}[5,9,12,14,15,3,6,14,11,1].\text{R5d1}[4,8,11,13,15,2,3,9,10,11].} \\ 
&\pmb{\text{R5d1}[3,7,10,13,14,6,4,11,14,13].\text{R5d1}[2,6,10,11,12,8,9,10,15,12].} \\ 
&\pmb{\text{R5d1}[1,6,7,8,9,13,14,12,1,2].\text{R5d1}[1,2,3,4,5,7,9,10,15,12]]}
\end{align*}
\end{tcolorbox}

\begin{tcolorbox}
    \begin{doublespace}
\noindent\(\{-5,-3,-4,-4,-6,-8,-2,5,-5,-3,-4,2,-3,8,-9\}\)
\end{doublespace}

\begin{doublespace}
\noindent\(\text{True}\)
\end{doublespace}
\end{tcolorbox}

\paragraph{{\bf Case 2 -}}
{\it {\bf 3-simplex operators} :}
The boxes below show the Mathematica codes for verifying if the 3-simplex operators in \eqref{eq:Rd3matrix-Case2} and \eqref{eq:Rd3matrix-Case2-BA} satisfy the spectral parameter dependent 3-simplex or tetrahedron equation. As mentioned in the Case 1 analysis, these operators depend on 3 parameters that are picked randomly from integers between -10 and 10. The output is shown in the second box.

\begin{tcolorbox}
    \begin{doublespace}
\noindent\(\pmb{\text{Ax}=\text{SparseArray}@A;}\\
\pmb{\text{sigmaZ}=\text{SparseArray}@\text{PauliMatrix}[3];}\\
\pmb{\mu=\text{Table}[\text{RandomInteger}[\{-10,10\}],\{i,1,15\}]}\\
\pmb{\text{R3dcase2}[\alpha \_,\beta \_,\gamma \_,\text{i$\_$},\text{j$\_$},\text{k$\_$}]\text{:=}} \\ \pmb{\mu[[i]]*\text{KroneckerProduct}\text{@@}\text{ReplacePart}[\text{ConstantArray}[\text{IdentityMatrix}[2],6],} \\ 
\pmb{\{\alpha
\to \text{Ax},\beta \to \text{Ax},\gamma \to \text{sigmaZ}\}]+}\\
\pmb{\mu[[j]]*\text{KroneckerProduct}\text{@@}\text{ReplacePart}[\text{ConstantArray}[\text{IdentityMatrix}[2],6],} \\ 
\pmb{\{\alpha \to \text{Ax},\beta \to
\text{sigmaZ},\gamma \to \text{Ax}\}] + }\\
\pmb{\mu[[k]]*\text{KroneckerProduct}\text{@@}\text{ReplacePart}[\text{ConstantArray}[\text{IdentityMatrix}[2],6],} \\ 
\pmb{\{\alpha \to \text{sigmaZ},\beta
\to \text{Ax},\gamma \to \text{Ax}\}];}\\
\pmb{\text{R3dcase21}[\alpha \_,\beta \_,\gamma \_,\text{i$\_$},\text{j$\_$},\text{k$\_$}]\text{:=}} \\ \pmb{\mu[[i]]*\text{KroneckerProduct}\text{@@}\text{ReplacePart}[\text{ConstantArray}[\text{IdentityMatrix}[2],6],} \\ 
\pmb{\{\alpha
\to \text{sigmaZ},\beta \to \text{sigmaZ},\gamma \to \text{Ax}\}]+}\\
\pmb{\mu[[j]]*\text{KroneckerProduct}\text{@@}\text{ReplacePart}[\text{ConstantArray}[\text{IdentityMatrix}[2],6],} \\ 
\pmb{\{\alpha \to \text{sigmaZ},\beta
\to \text{Ax},\gamma \to \text{sigmaZ}\}] + }\\
\pmb{\mu[[k]]*\text{KroneckerProduct}\text{@@}\text{ReplacePart}[\text{ConstantArray}[\text{IdentityMatrix}[2],6],} \\ 
\pmb{\{\alpha \to \text{Ax},\beta \to
\text{sigmaZ},\gamma \to \text{sigmaZ}\}];}\\
\pmb{\text{Simplify}[\text{R3dcase2}[1,2,3,8,11,12].\text{R3dcase2}[1,4,5,3,7,9].\text{R3dcase2}[2,4,6,13,10,8].} \\ 
\pmb{\text{R3dcase2}[3,5,6,14,10,12]\text{==}}\\
\pmb{\text{R3dcase2}[3,5,6,14,10,12].\text{R3dcase2}[2,4,6,13,10,8].\text{R3dcase2}[1,4,5,3,7,9].} \\ 
\pmb{\text{R3dcase2}[1,2,3,8,11,12]]}\\
\pmb{\text{Simplify}[\text{R3dcase21}[1,2,3,8,11,12].\text{R3dcase21}[1,4,5,3,7,9].} \\ 
\pmb{\text{R3dcase21}[2,4,6,13,10,8].\text{R3dcase21}[3,5,6,14,10,12]\text{==}}\\
\pmb{\text{R3dcase21}[3,5,6,14,10,12].\text{R3dcase21}[2,4,6,13,10,8].} \\ 
\pmb{\text{R3dcase21}[1,4,5,3,7,9].\text{R3dcase21}[1,2,3,8,11,12]]}\)
\end{doublespace}
\end{tcolorbox}

\begin{tcolorbox}
    \begin{doublespace}
\noindent\(\{-2,10,-4,-3,-2,-4,0,-10,-5,-5,7,10,-5,4,9\}\)
\end{doublespace}

\begin{doublespace}
\noindent\(\text{True}\)
\end{doublespace}

\begin{doublespace}
\noindent\(\text{True}\)
\end{doublespace}

\end{tcolorbox}

{\it {\bf 4-simplex operators} :} The Mathematica code for a linear combination of the $(2,2)$ and $(4,0)$ types 4-simplex operator in \eqref{eq:R1d4matrix-Case2} and \eqref{eq:R2d4matrix-Case2noninvertible} is shown in the box below. For a random set of integer coefficients, this operator satisfies the spectral parameter dependent 4-simplex equation. This is seen in the output box below.

\begin{tcolorbox}
    \begin{align*}
&\pmb{\text{Ax}=\text{SparseArray}@A;}\\
&\pmb{\text{sigmaZ}=\text{SparseArray}@\text{PauliMatrix}[3];}\\
&\pmb{\mu=\text{Table}[\text{RandomInteger}[\{-10,10\}],\{i,1,15\}]}\\
&\pmb{\text{R4d}[\alpha \_,\beta \_,\gamma \_,\delta \_,\text{i$\_$},\text{j$\_$},\text{k$\_$},\text{l$\_$},\text{m$\_$},\text{n$\_$},\text{p$\_$},\text{q$\_$}]\text{:=}}\\
&\pmb{\mu[[i]]*\text{KroneckerProduct}\text{@@}\text{ReplacePart}[\text{ConstantArray}[\text{IdentityMatrix}[2],10],} \\ 
&\pmb{\{\alpha \to \text{Ax},\beta \to
\text{Ax},\gamma \to \text{sigmaZ},\delta \to \text{sigmaZ}\}]+}\\
&\pmb{\mu[[j]]*\text{KroneckerProduct}\text{@@}\text{ReplacePart}[\text{ConstantArray}[\text{IdentityMatrix}[2],10],} \\ 
&\pmb{\{\alpha \to \text{Ax},\beta \to
\text{sigmaZ},\gamma \to \text{Ax},\delta \to \text{sigmaZ}\}]+}\\
&\pmb{\mu[[k]]*\text{KroneckerProduct}\text{@@}\text{ReplacePart}[\text{ConstantArray}[\text{IdentityMatrix}[2],10],} \\ 
&\pmb{\{\alpha \to \text{sigmaZ},\beta
\to \text{Ax},\gamma \to \text{Ax},\delta \to \text{sigmaZ}\}]+}\\
&\pmb{\mu[[l]]*\text{KroneckerProduct}\text{@@}\text{ReplacePart}[\text{ConstantArray}[\text{IdentityMatrix}[2],10],} \\ 
&\pmb{\{\alpha \to \text{Ax},\beta \to
\text{sigmaZ},\gamma \to \text{sigmaZ},\delta \to \text{Ax}\}] +}\\
&\pmb{ \mu[[m]]*\text{KroneckerProduct}\text{@@}\text{ReplacePart}[\text{ConstantArray}[\text{IdentityMatrix}[2],10],} \\ 
&\pmb{\{\alpha \to \text{sigmaZ},\beta
\to \text{Ax},\gamma \to \text{sigmaZ},\delta \to \text{Ax}\}] +}\\
&\pmb{\mu[[n]]*\text{KroneckerProduct}\text{@@}\text{ReplacePart}[\text{ConstantArray}[\text{IdentityMatrix}[2],10],} \\ 
&\pmb{\{\alpha \to \text{sigmaZ},\beta
\to \text{sigmaZ},\gamma \to \text{Ax},\delta \to \text{Ax}\}] +}\\
&\pmb{\mu[[p]]*\text{KroneckerProduct}\text{@@}\text{ReplacePart}[\text{ConstantArray}[\text{IdentityMatrix}[2],10],} \\ 
&\pmb{\{\alpha \to \text{Ax},\beta \to
\text{Ax},\gamma \to \text{Ax},\delta \to \text{Ax}\}]+}\\
&\pmb{\mu[[q]]*\text{KroneckerProduct}\text{@@}\text{ReplacePart}[\text{ConstantArray}[\text{IdentityMatrix}[2],10],} \\ 
&\pmb{\{\alpha \to \text{sigmaZ},\beta
\to \text{sigmaZ},\gamma \to \text{sigmaZ},\delta \to \text{sigmaZ}\}];}\\
&\pmb{\text{Simplify}[\text{R4d}[1,2,3,4,11,12,8,7,9,10,5,13].\text{R4d}[1,5,6,7,11,15,2,5,6,7,8,14].} \\ 
&\pmb{\text{R4d}[2,5,8,9,12,15,8,13,11,15,12,10].}\\
&\pmb{\text{R4d}[3,6,8,10,13,6,1,4,14,9,5,7].\text{R4d}[4,7,9,10,14,7,2,5,8,10,11,13]==}\\
&\pmb{\text{R4d}[4,7,9,10,14,7,2,5,8,10,11,13].\text{R4d}[3,6,8,10,13,6,1,4,14,9,5,7].} \\ 
&\pmb{\text{R4d}[2,5,8,9,12,15,8,13,11,15,12,10].}\\
&\pmb{\text{R4d}[1,5,6,7,11,15,2,5,6,7,8,14].\text{R4d}[1,2,3,4,11,12,8,7,9,10,5,13]]}
\end{align*}
\end{tcolorbox}

\begin{tcolorbox}
\begin{doublespace}
\{-10,-7,9,10,10,-3,-1,10,10,5,6,-7,9,-5,10\} \\
\text{True}
\end{doublespace}
\end{tcolorbox}

{\it {\bf 5-simplex operators} :} The Mathematica code showing that the $(4,1)$ type 5-simplex operator satisfies the 5-simplex equation is shown below.
\begin{tcolorbox}
  \begin{align*}
&\pmb{\text{Ax}=\text{SparseArray}@A;}\\
&\pmb{\text{sigmaZ}=\text{SparseArray}@\text{PauliMatrix}[3];}\\
&\pmb{\mu=\text{Table}[\text{RandomInteger}[\{-10,10\}],\{i,1,15\}]}\\
&\pmb{\text{R5d2}[\alpha \_,\beta \_,\gamma \_,\delta \_,\eta \_,\text{i$\_$},\text{j$\_$},\text{k$\_$},\text{l$\_$},\text{m$\_$}]\text{:=}}\\
&\pmb{\mu[[i]]*\text{KroneckerProduct}\text{@@}\text{ReplacePart}[\text{ConstantArray}[\text{IdentityMatrix}[2],15],} \\ 
&\pmb{\{\alpha \to \text{Ax},\beta \to
\text{Ax},\gamma \to \text{Ax},\delta \to \text{Ax},\eta \to \text{sigmaZ}\}]+}\\
&\pmb{\mu[[j]]*\text{KroneckerProduct}\text{@@}\text{ReplacePart}[\text{ConstantArray}[\text{IdentityMatrix}[2],15],} \\ 
&\pmb{\{\alpha \to \text{Ax},\beta \to
\text{Ax},\gamma \to \text{Ax},\delta \to \text{sigmaZ},\eta \to \text{Ax}\}]+}\\
&\pmb{\mu[[k]]*\text{KroneckerProduct}\text{@@}\text{ReplacePart}[\text{ConstantArray}[\text{IdentityMatrix}[2],15],} \\ 
&\pmb{\{\alpha \to \text{Ax},\beta \to
\text{Ax},\gamma \to \text{sigmaZ},\delta \to \text{Ax},\eta \to \text{Ax}\}]+}\\
&\pmb{\mu[[l]]*\text{KroneckerProduct}\text{@@}\text{ReplacePart}[\text{ConstantArray}[\text{IdentityMatrix}[2],15],} \\ 
&\pmb{\{\alpha \to \text{Ax},\beta \to
\text{sigmaZ},\gamma \to \text{Ax},\delta \to \text{Ax},\eta \to \text{Ax}\}] +}\\
&\pmb{\mu[[m]]*\text{KroneckerProduct}\text{@@}\text{ReplacePart}[\text{ConstantArray}[\text{IdentityMatrix}[2],15],} \\ 
&\pmb{\{\alpha \to \text{sigmaZ},\beta
\to \text{Ax},\gamma \to \text{Ax},\delta \to \text{Ax},\eta \to \text{Ax}\}];}\\
&\pmb{\text{Simplify}[\text{R5d2}[1,2,3,4,5,7,9,10,15,12].\text{R5d2}[1,6,7,8,9,13,14,12,1,2].} \\
&\pmb{\text{R5d2}[2,6,10,11,12,8,9,10,15,12].\text{R5d2}[3,7,10,13,14,6,4,11,14,13].} \\ 
&\pmb{\text{R5d2}[4,8,11,13,15,2,3,9,10,11].\text{R5d2}[5,9,12,14,15,3,6,14,11,1]\text{==}}\\
&\pmb{\text{R5d2}[5,9,12,14,15,3,6,14,11,1].\text{R5d2}[4,8,11,13,15,2,3,9,10,11].} \\ 
&\pmb{\text{R5d2}[3,7,10,13,14,6,4,11,14,13].\text{R5d2}[2,6,10,11,12,8,9,10,15,12].} \\ 
&\pmb{\text{R5d2}[1,6,7,8,9,13,14,12,1,2].\text{R5d2}[1,2,3,4,5,7,9,10,15,12]]}
\end{align*}
\end{tcolorbox}

\begin{tcolorbox}
\begin{doublespace}
\noindent\(\{4,-4,-5,-5,-3,-4,6,-4,1,-5,4,-5,1,5,-6\}\)
\end{doublespace}

\begin{doublespace}
\noindent\(\text{True}\)
\end{doublespace}

\end{tcolorbox}

\bibliographystyle{acm}
\normalem
\bibliography{refs}

\end{document}